\documentclass[12pt,preprint]{aastex}

\def\spose#1{\hbox to 0pt{#1\hss}}
\def\approxlt{\mathrel{\spose{\lower 3pt\hbox{$\sim$}}
        \raise 2.0pt\hbox{$<$}}}
\def\approxgt{\mathrel{\spose{\lower 3pt\hbox{$\sim$}}
        \raise 2.0pt\hbox{$>$}}}

\def\multleft#1{\hbox to size{\vbox {\halign {\lft{##}\cr #1}}\hfill}\par}
\def\multright#1{\hbox to size{\vbox {\halign {\rt{##}\cr #1}}\hfill}\par}

\def\boxit#1{\vbox{\hrule\hbox{\vrule\kern3pt\vbox{\kern3pt
          #1 \kern3pt}\kern3pt\vrule}\hrule}}

\def\cm{{\rm\thinspace cm}}

\def\erg{{\rm\thinspace erg}}

\def\Msun{\hbox{$\rm\thinspace M_{\odot}$}}

\def\s{{\rm\thinspace s}}

\def\ergpcmsqps{\hbox{$\erg\cm^{-2}\s^{-1}\,$}}

\def\ergps{\hbox{$\erg\s^{-1}\,$}}

\def\L{\hbox{$\mathcal L$}}

\received{}
\accepted{}

\slugcomment{submitted to ApJ} 

\shorttitle{Spatial correlation function of {\it Chandra} Selected AGNs}
\shortauthors{Yang et al.}

\begin{document}

\title{Spatial Correlation Function of the {\it Chandra} Selected Active Galactic Nuclei}

\author{
Y. Yang\altaffilmark{1,2}, 
R. F. Mushotzky\altaffilmark{2}, 
A. J. Barger\altaffilmark{3,4}, 
L. L. Cowie\altaffilmark{4}, 
}

\altaffiltext{1}{Department of Astronomy, University of Maryland, College Park 20742}
\altaffiltext{2}{Laboratory for High Energy Astrophysics, Goddard Space Flight Center, Code 660, NASA, Greenbelt, MD, 20770}
\altaffiltext{3}{Department of Astronomy, University of Wisconsin, Madison, WI 53760} 
\altaffiltext{4}{Institute for Astronomy, University of Hawaii, Honolulu, HI 96822} 

\begin{abstract}
We present the spatial correlation function analysis of non-stellar X-ray point sources
in the {\it Chandra} Large Area Synoptic X-ray Survey of Lockman 
Hole Northwest (CLASXS).  Our 9 ACIS-I fields cover a contiguous solid 
angle of 0.4 deg$^{2}$ and reach a depth of   
$3 \times 10^{-15}$~\ergpcmsqps in the 2--8 keV band. We supplement our analysis
with data from the Chandra Deep Field North (CDFN). The addition of this field
allows better probe of the correlation function at small scales. A total of 233 and 
252 sources with spectroscopic information are used in the study of
the CLASXS and CDFN fields respectively.

We calculate both 
redshift-space and projected correlation functions in comoving coordinates,
averaged over the redshift range of $0.1<z<3.0$, for both CLASXS and CDFN fields
for a standard cosmology with $\Omega_{\Lambda} = 0.73, \Omega_{M} = 0.27$, and $h = 0.71$
($H_0 = 100h$~km~s$^{-1}$~Mpc$^{-1}$). 
The correlation function for the CLASXS field over scales of 3~Mpc$<s<$~200~Mpc 
can be modeled as a power-law of the form $\xi(s) = (s/s_0)^{-\gamma}$, 
with $\gamma = 1.6^{+0.4}_{-0.3}$ and $s_0 = 8.0^{+1.4}_{-1.5}$~Mpc. 
The redshift-space correlation function for CDFN on scales of 1~Mpc$<s<$~100~Mpc is 
found to have a similar correlation length $s_0 = 8.55^{+0.75}_{-0.74}$~Mpc, 
but a shallower slope ($\gamma = 1.3 \pm 0.1$). The real-space correlation 
functions derived from the projected correlation functions, are found to be 
$r_0 = 8.1^{+1.2}_{-2.2}$~Mpc, and $\gamma = 2.1 \pm 0.5$ for the CLASXS field, 
and $r_0 = 5.8^{+1.0}_{-1.5}$~Mpc, $\gamma = 1.38^{+0.12}_{-0.14}$ for the CDFN field. 
By comparing the real- and redshift-space correlation functions in the combined CLASXS and 
CDFN samples, we are able to estimate the redshift distortion parameter $\beta = 0.4 \pm 0.2$ at 
an effective redshift $z = 0.94$. We compare the correlation functions for hard and soft 
spectra sources in the CLASXS field and find no significant difference between the two groups. 
We have also found that the correlation between  X-ray luminosity and
clustering amplitude is weak, which, however, is fully consistent with
the expectation using the simplest relations between X-ray luminosity, 
blackhole mass, and dark halo mass. 

We study the evolution of the AGN clustering by dividing the samples into 4 redshift bins over
0.1~Mpc$<z<$3.0~Mpc. We find a very mild evolution in the clustering amplitude, which show the same
evolution trend found in optically selected quasars in the 2dF survey.  We estimate the 
evolution of the bias, and find that the bias increases rapidly with redshift ($b(z=0.45) = 0.95 \pm 0.15$and $b(z = 2.07)= 3.03 \pm 0.83$). The typical mass of the dark matter halo derived from the bias estimates
show little change with redshift. The average  halo mass is found to be  $\log~(M_{halo}/M_{\sun}) \sim 12.1$.  
\end{abstract}


\section{Introduction}

Structure formation and evolution in the universe and the 
formation and growth of supermassive black holes (SMBHs) are 
two fundamental problems in astronomy which are still not fully 
understood. While recent progresses in the cosmic microwave background, 
the high redshift type Ia supernovae survey, and the large optical surveys 
have significantly improved our understanding of the evolution of large scale structure,
there are still several gaps in the picture of structure formation. The data at redshift 
of $\sim 1$, where most of the cosmic star formation might
have taken place, is still very limited. On scales of galaxies and cluster of galaxies,
the feed back process from galaxies or AGNs could significantly alter structure formation
models where gravitation is the only driving force. The clustering of active galactic 
nuclei (AGNs) provides unique path to the solution of these problems because (1) the 
AGNs are often bright compared to normal galaxies and are easily 
seen at large cosmological distance; (2) AGNs trace the violent growth 
phase of SMBHs and hence their clustering properties provide a link 
between the dark matter halo to the AGN activity. Large scale AGN 
surveys have been traditionally carried out in the optical band with 
dedicated telescopes. The most recent of these are the
Sloan Digital Sky Survey (SDSS, {Schneider} {et~al.} 2004) and the Two 
Degree Field Survey (2dF, {Croom} {et~al.} 2005, C05 hereafter).  These surveys 
have demonstrated that the clustering of AGNs can be used to measure cosmological 
parameters ({Croom} {et~al.} 2004; {Outram} {et~al.} 2004), and to constrain gravitational 
lensing ({Myers} {et~al.} 2003).  However, as has been found in recent 
{\it Chandra} and {\it XMM-Newton} deep surveys, a large fraction of X-ray 
detected AGNs show little or no activity in optical observations (e.g. {Barger} {et~al.} 2005), 
most probably due to obscuration ({Fabian} \& {Iwasawa} 1999) but possibly with some contribution from 
light dilution of the host galaxy ({Moran}, {Filippenko}, \& {Chornock} 2002), though Barger et al. 
(2005) argue this is not a dominant factor. 
This results in a large fraction of AGNs being missed in the optical surveys. On the other hand, 
hard X-rays ($>2$~keV) is almost unaffected 
by obscuring column densities $N_{H}< 10^{24}$~cm$^{-2}$, and the X-ray 
emission from the host galaxies is low compared to the AGNs. Thus hard X-rays are  
at present the best energy band to find AGNs ({Mushotzky} 2004).  Recent optical follow-ups 
of {\it Chandra} deep surveys have revealed that the hard  X-ray sources 
are mostly found around $z \sim 1$ instead of $z \sim 2$ as seen in optically 
selected quasar samples. The low redshift population is dominated by non-broadline 
objects, while broadline AGNs are found mostly at higher redshifts ({Steffen} {et~al.} 2003). 
Given these new discoveries, it is important to know how 
the clustering properties of X-ray and optical selected AGN differ. 

The most extensive X-ray AGN surveys so far performed used the 
{\it ROSAT} telescope ({Mullis} {et~al.} 2004). Because the telescope 
is not sensitive above 2 keV, {\it ROSAT} misses a large fraction of hard X-ray sources . 
The relatively poor spatial resolution of ROSAT also limits the accuracy of optical 
identifications. Most of the {\it ROSAT} detected AGNs show broad emission lines in their
 optical spectra.  Both the optical quasar surveys and the ROSAT 
surveys suggest that AGNs have correlation properties similar to 
the local galaxies. The results seem to be independant of the sample medium redshifts.
This result is puzzling because AGNs are believed to be preferentially form in high density peaks where interactions between
galaxies are more common, and interactions in turn are thought to be crucial in AGN 
fueling ({Di Matteo}, {Springel}, \&  {Hernquist} 2005). The mass of the dark matter
halos that host AGNs are hence likely to be more massive.   

The clustering results on hard X-ray AGNs are so far contradictory. 
Earlier studies of a small number of individual {\it Chandra} fields
seem to indicate that the hard band number counts in these 
small fields has fluctuations larger than expected from Poisson noise 
({Cowie} {et~al.} 2002; {Manners} {et~al.} 2003) but the result is contradicted 
with larger samples of {\it Chandra} fields ({Kim} {et~al.} 2004).
{Basilakos} {et~al.} (2004) found a 4$\sigma$ clustering signal in hard X-ray  
sources at $f_{2-8 keV} > 10^{-14}$~\ergpcmsqps using angular 
correlation functions on a XMM detected AGN sample from a 2 deg$^{2}$ survey. 
A similar result was also found earlier in our 0.4~deg$^{2}$ {\it Chandra} field 
(see below) using the count-in-cells technique ({Yang} {et~al.} 2003). Using the Limber equation  
Basilacos et al. (2004) argue that the hard X-ray sources are likely to be more 
strongly clustered than the optically selected AGNs. {Gilli} {et~al.} (2003) reported the detection of 
large angular-redshift clustering in the Chandra Deep field South, which 
seems to be dominated by hard X-ray sources. Using the projected correlation
function for the optically identified X-ray sources from the Chandra Deep 
field North (CDFN) and South (CDFS), {Gilli} {et~al.} (2005) found that the 
average correlation amplitude in the CDFS is higher than that in the CDFN, and the 
latter is consistent with the correlation amplitude found in optically detected 
quasars.    

In this paper, we report our spatial correlation function analysis of 
the optically identified X-ray sources in the {\it Chandra} 
Large Area Synoptic X-ray Survey of the ISO Lockman Hole Northwest region 
(CLASXS). CLASXS is so far the largest contiguous {\it Chandra} deep
field with a high level of spectroscopic identifications. The size of 
the field is chosen to reduce the cosmic variance in the X-ray background 
to $\sim 10\%$ (Yang et al. 2004). For comparison, we have also analyzed the correlation
functions for the CDFN field, using the published
X-ray catalog by {Alexander} {et~al.} (2003) and the optical catalog of 
{Barger} {et~al.} (2003). 
Because the two surveys use basically the same optical instruments in the 
follow-up observations, and thus have the same accuracy in redshift measurements,
the comparison is relatively straight forward. The LogN-LogS of the CDFN 
agrees well with that of CLASXS, also indicating that the CDFN is 
a ``typical'' field. 
The depth of the CDFN is very useful in probing the correlation function at small 
separations. In \S~\ref{sec_obs_data} we summarize
our observations and data analysis. In \S~\ref{sec_methods} we discuss the methodology we use in 
the clustering analysis. The results of the correlation functions
are presented in \S~\ref{sec_results}. The evolution of AGN clustering is presented in \S~\ref{sec_evolution}. 
In \S~\ref{sec_discussion} we discussion the implications of our results. Finally summarize our results in \S~\ref{sec_conclusion}. Throughout this paper, unless noted otherwise, 
we assume $H_{0}=71$ and a flat universe with $\Omega_{M}=0.27$ and  
$\Omega_{\Lambda} = 0.73$. 
 
\section{Observations and data
\label{sec_obs_data}}   
CLASXS is a 0.4~deg$^{2}$ contiguous field centered at 
$\alpha=10^{h}34^{m}$, $\delta=57\arcdeg40\arcmin$ (J2000) in 
the very low galactic absorption Lockman Hole Northwest region.
It is the deepest 170$\mu$m survey field observed by ISOPHOT instrument on board
{\it ISO}, and has recently been observed by the {\it Spitzer Space telescope} 
({Lonsdale} {et~al.} 2004).  The {\it Chandra} observation consists 
of 9 ACIS-I pointings separated from each other by 
$\sim 10\arcmin$ to allow a close to uniform sky coverage. The center field
has an exposure time of $\sim 70$~ks while the other eight pointings, have exposure 
times of $\sim 40$~ks. The exposure time were designed to give a uniform flux limit
of $f_{2-8~keV} \approx 5 \times 10^{-15}$~\ergpcmsqps, which is about
a factor of 2 below the ``knee'' of the LogN-LogS curve. This choice
of sensitivity allows a proper sampling of the X-ray background sources and 
also achieves a highest source finding efficiency.

\begin{figure}
\includegraphics[scale=.7,angle=90]{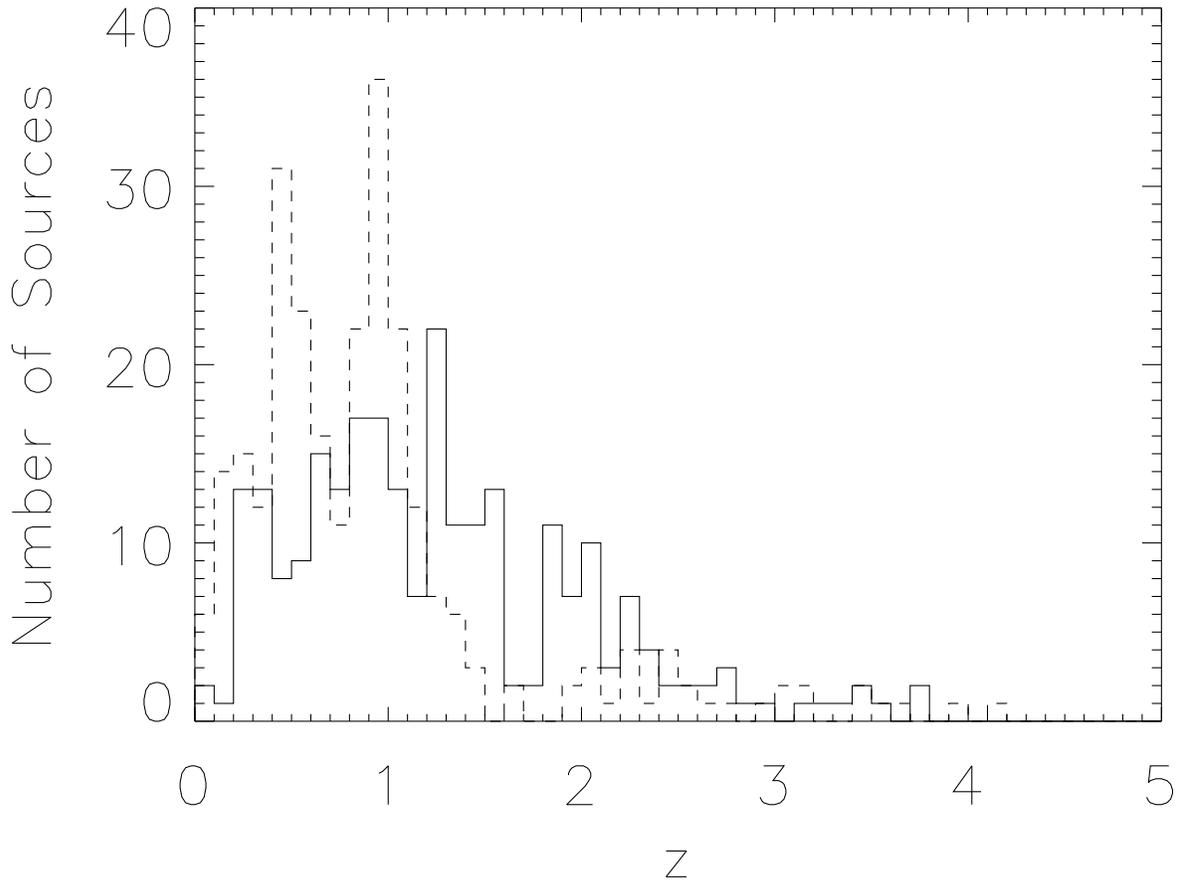}
\caption[z_dist.ps] {Redshift distribution of the optically identified 
X-ray sources. Solid line: the CLASXS field; dashed line: the CDFN field.
\label{z_dist}}
\end{figure}

\begin{figure}
\includegraphics[scale=.7,angle=90]{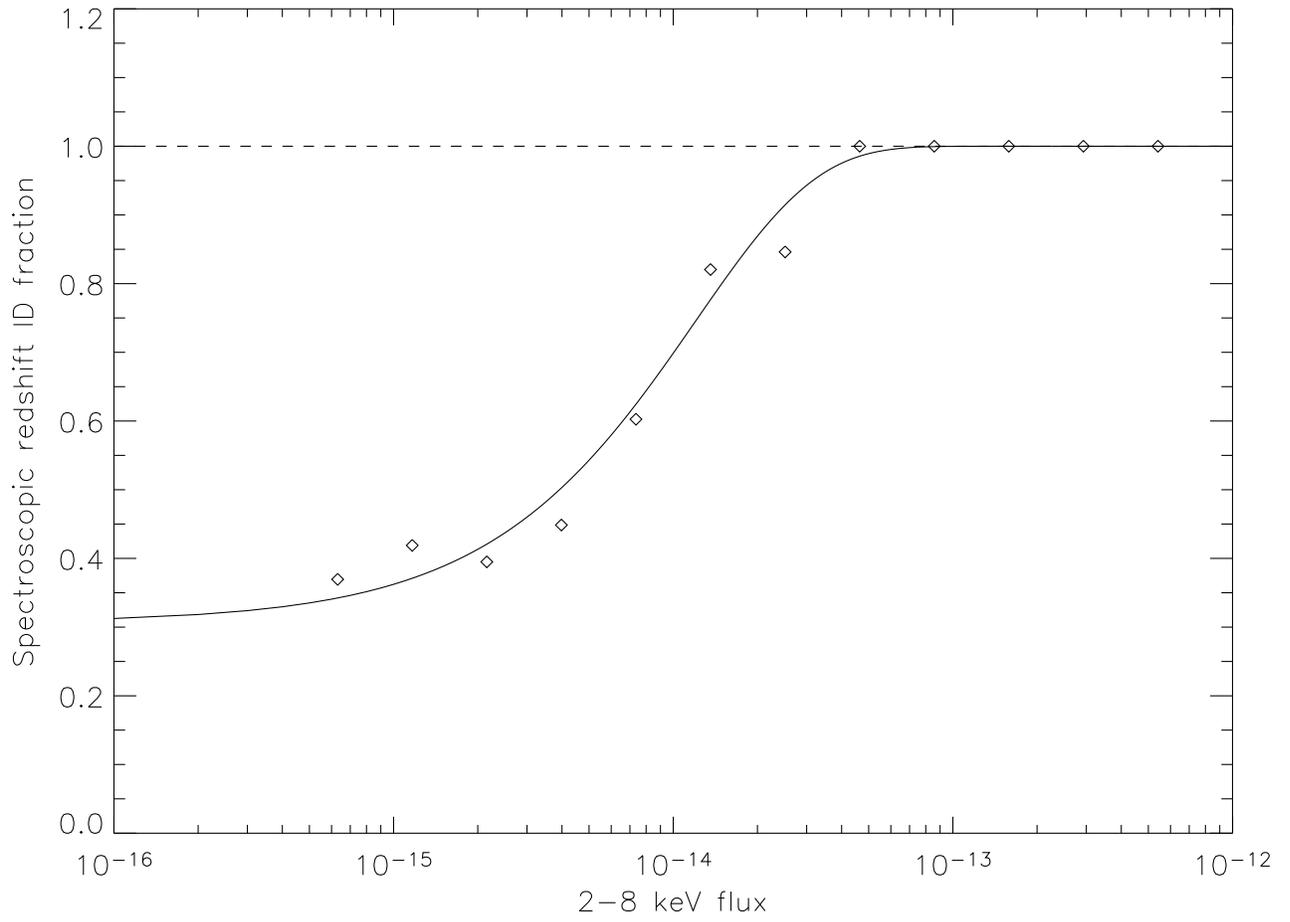}
\caption[fig2.ps] {The optical identification fraction as a function of 
2--8 keV flux. Solid line shows the best-fit.
\label{optical_id}}
\end{figure}

The sub-arcsecond spatial resolution of {\it Chandra} 
observatory allows an unambiguous optical identification of the X-ray sources, particularly, for 
those which appear to be normal galaxies in optical band. Combined with follow-ups using 
the large Keck and Subaru optical telescopes, we identified and measured the redshifts 
in a large fraction of the X-ray detected AGNs in our survey. The details 
and the catalogs of the survey can be found in {Yang} {et~al.} (2004) and 
{Steffen} {et~al.} (2004). We performed spectroscopic 
observations for $\sim 90\%$ of the 525 detected X-ray sources. A total of 272 spectroscopic 
redshifts have been obtained, while the spectra of the rest of the sources have a signal-to-noise ratio
too low to obtain secure redshift measurements. The redshift distribution of the identified sources 
are shown in Figure~\ref{z_dist}. The fraction of sources with spectroscopic redshift as a function of 
hard X-ray flux is shown in Figure~\ref{optical_id}. 

The 2~Ms CDFN is so far the deepest {\it Chandra} field, reaching a flux limit
of $f_{2-8keV} \approx 1.4 \times 10^{-16}$~\ergpcmsqps ({Alexander} {et~al.} 2003).
This is $\sim 20$ times deeper than the CLASXS field. The areal density of sources
in CDFN is also $\sim 5$ times higher. The optical observation were performed using the same
telescope as CLASXS ({Barger} {et~al.} 2003). We use the published catalog, which 
contains 306 sources with spectroscopic redshift. The redshift distribution of the CDFN 
sources is also shown in Figure~\ref{z_dist}. The fainter X-ray sources in the CDFN are 
more likely to be found at low redshift, $z<1$, compared to the CLASXS sources.        

\section{Methods
\label{sec_methods}}

To quantify spatial clustering in a point process, the most commonly
used technique is the two point correlation function. In short, a two point 
correlation function measures the excess probability of finding a pair of objects
as a function of pair separation ({Peebles} 1980).  
\begin{equation}
dP = n_{0}^{2}[1+\xi(r)]dV_{1}dV_{2}
\label{acf_def}
\end{equation}
where $n_{0}$ is the mean density and $r$ is the {\it comoving} distance
between two sources.

Observations of low redshift galaxies and clusters of 
galaxies show that the correlation function of these objects 
over a wide range of scales can be described by a 
power-law
\begin{equation}
\xi(r) =(\frac{r}{r_{0}})^{-\gamma},
\label{powerlaw}
\end{equation}
with $\gamma \sim 1.6-1.9$ ({Peebles} 1980; {Peacock} 1999).  It should be noted that the correlation
function is in fact a function of redshift, which we will discuss in 
\S~\ref{sec_evolution}. 
Because of the small sample sizes of most of the AGN surveys, 
correlation functions over very wide redshift ranges are commonly
used. This only makes sense if the clustering is almost constant in
comoving coordinates. Fortunately, this is very close to the truth, 
as we shall see in \S~\ref{sec_evolution}. 

\subsection{Redshift- and real-space Correlation functions
\label{subsec_corr_func}}
The nominal distance between sources calculated using the sky coordinates of the 
sources and their redshifts is sometimes called distance
in {\it redshift-space}, we shall use $\bf{s}$ instead of $\bf{r}$ 
to indicate the distance calculated this way. 
It is apparent that the line-of-sight peculiar velocity of the sources 
could also contribute to the measured redshift (redshift distortion). 
This effect is most important at separations smaller than the correlation 
length.  The projected correlation function, which computes 
the integrated correlation function along the line-of-sight and is not 
affected by redshift distortion, is often used to obtain the real-space 
correlation function ({Peebles} 1980). The projection, however, could make the 
correlation signal more difficult to measure. In small fields like the CDFN, 
the projected correlation function is also restricted
by the field size, and could be affected by cosmic variance. We will calculate 
both the redshift-space and projected correlation functions 
in this paper. This allows us to estimate the effects of redshift distortion. 

Following {Davis} \& {Peebles} 1983, we define $\bf v_{1}$ and $\bf v_{2}$
to be the positions of two sources in the redshift-space, ${\bf s} \equiv 
{\bf v_{1}} - {\bf v_{2}}$ to be the redshift-space separation, and 
 ${\bf l} \equiv ( {\bf v_{1}} + {\bf v_{2}} )/2$ to be the mean distance to 
the pair of sources. We can then compute the correlation function $\xi(r_{p},\pi)$ on 
a two dimensional grid, where $\pi$ and $r_{p}$ are separations along
and across the line-of-sight respectively:  
\begin{equation}
\pi=\frac{{\bf s} \cdot {\bf l}}{{\bf |l|}},
\end{equation}
\begin{equation}
r_p=\sqrt{{\bf s} \cdot {\bf s} - {\bf \pi}^2}.
\end{equation}
The projected correlation function is defined as the line-of-sight
integration of $\xi(r_{p},\pi)$:
\begin{equation}
 w_p(r_p)= \int_{-\pi_{max}}^{\pi_{max}} d\pi \ \xi(r_p,\pi)= \int_{-\pi_{max}}^{\pi_{max}}
dy \ \xi(\sqrt{r_p^2+y^2}),
\label{wp}
\end{equation}
where $y$ is the line-of-sight separation. It has been shown 
({Davis} \& {Peebles} 1983) that, when $\pi_{max} \rightarrow \infty$, 
$w_p(r_p)$ satisfies a simple relation
with the real-space correlation function. If a power-law form in Equation
\ref{powerlaw} is assumed, then
 \begin{equation}
w_p(r_p)=r_p \left(\frac{r_0}{r_p}\right)^\gamma
\frac{\Gamma(\frac{1}{2})\Gamma(\frac{\gamma-1}{2})}{\Gamma(\frac{\gamma}{2})}.
\label{wp_n_real}
\end{equation}
In practice, the integration is not performed to very large separations because 
the major contribution to the projected signal comes from separations of a few times the correlation 
length $s_{0}$. Integrating to larger $\pi$ will only add noise to the results.  
After testing various scales, we found  $\pi_{max} = 20 - 40$~Mpc produces
consistent results for our samples. 

\subsection{Correlation function Estimator}  
To obtain an unbiased estimate of the correlation function, we must correct for selection
effects. Usually, these selection effects are treated using
random samples generated with computer simulations. By comparing
the simulated and observed pair distributions, the selection functions 
effectively cancel. 
We compute the correlation function using the minimum variance estimator 
\begin{equation}
\xi = \frac{DD-2DR+RR}{RR}
\end{equation}
where $DD$, $DR$ and $RR$ are the numbers of data-data, data-random and random-random
pairs respectively, with comoving distances $s_{0}-\Delta s/2 < s < s_{0} + \Delta s/2$
(L-S estimator, {Landy} \& {Szalay} 1993). The random catalog is produced through simulations described below to 
account for the selection effects in observations. The random catalog 
usually contains a very large number of objects so that the Poisson
noise introduced is negligible. We have checked our results using both L-S and the 
Davis-Peebles estimators ({Davis} \& {Peebles} 1983) and found very good agreement
between the two methods.

\subsection{Uncertainties of correlation functions}
There are two terms in the uncertainty of the correlation function: the statistical 
fluctuations and the cosmic variance. The statistical uncertainty of the correlation 
function is estimated assuming the error of the DR and RR pairs are zero, and the 
uncertainty of DD is Poissonian, 
\begin{equation}
\sigma_{\xi} = \frac{(1+\xi)}{\sqrt{DD}}
\label{Poiss}
\end{equation}
In the case of small DD, where $\sqrt{DD}$ underestimates the error, we use the 
approximation formula ({Gehrels} 1986) to calculate the Poisson upper and lower limits. 
Since the DDs are in fact correlated, the use of Poisson errors could underestimate
the real uncertainty.  In the literature {\it bootstrap} resampling ({Efron} 1982) is often 
used to calculate the errors of the correlation function. The method
is particularly useful in cases when the probability 
distribution function (PDF) of the variable is unknown, or in cases 
when the variables are derived from  Poissonian distributed data using 
complex transformations, which results in rather complex PDFs.  
{Mo}, {Jing}, \& {Boerner} (1992) showed that in the case of large DD, 
the bootstrap error is $\sim \sqrt 3$ of the Poisson error. 
We use Poisson errors in our redshift-space correlation function
estimates. On the other hand, we use bootstrap methods when estimating 
the uncertainties of the projected correlation function. This is because
the numerical integration used in Equation~\ref{wp} make it difficult to 
apply Poisson errors directly.

Cosmic variance is known to affect the estimation of the mean density  
when applied to small samples of normal galaxies from optical surveys.
Such effect, however, is likely to be small on our X-ray selected AGN
sample for the following reasons. The volumn
of our survey is very large compared to the typical pensil-beam optical survey of
normal galaxies that typically covers very narrow redshift ranges.
On the other hand, the number density of AGNs is much lower than
that of the normal galaxies, making it hard to trace individual
structures at high enough sampling rate. The window function of the spectroscopic
follow-up in our survey is also very flat over a wide redshift range 
(except in the redshift desert at $z \sim 1.2-2$). 
The combination of these factors makes it very difficult for a 
small number of structures been over sampled and thus producing 
incorrect estimation of the mean density.  However, for 
ultra deep surveys with field size of a single {\it Chandra} field, 
small number of velocity spikes can indeed affect the correlation analysis,
as seen in the case of the Chandra Deep Field South. Such
structure, however, will affect number counts in the field at flux
levels comparable to the depth of CLASXS. 
Based on the very good agreement among the number counts found in the CLASXS, 
CDFN, and other deep surveys (Yang et al. 2004), we believe the uncertainty from 
cosmic variance on the whole sample is likely to be small. 
However, the cosmic variance effect on subsamples could still be important, as seen in 
\S \ref{sec_evolution}. In such cases, using statistical uncertainty alone could 
underestimates the true uncertainty.      

\subsection{The mock catalog}
To account for the observational selection and edge effects, we perform 
extensive simulations to construct a mock catalog. 

\begin{figure}
\includegraphics[scale=.7,angle=90]{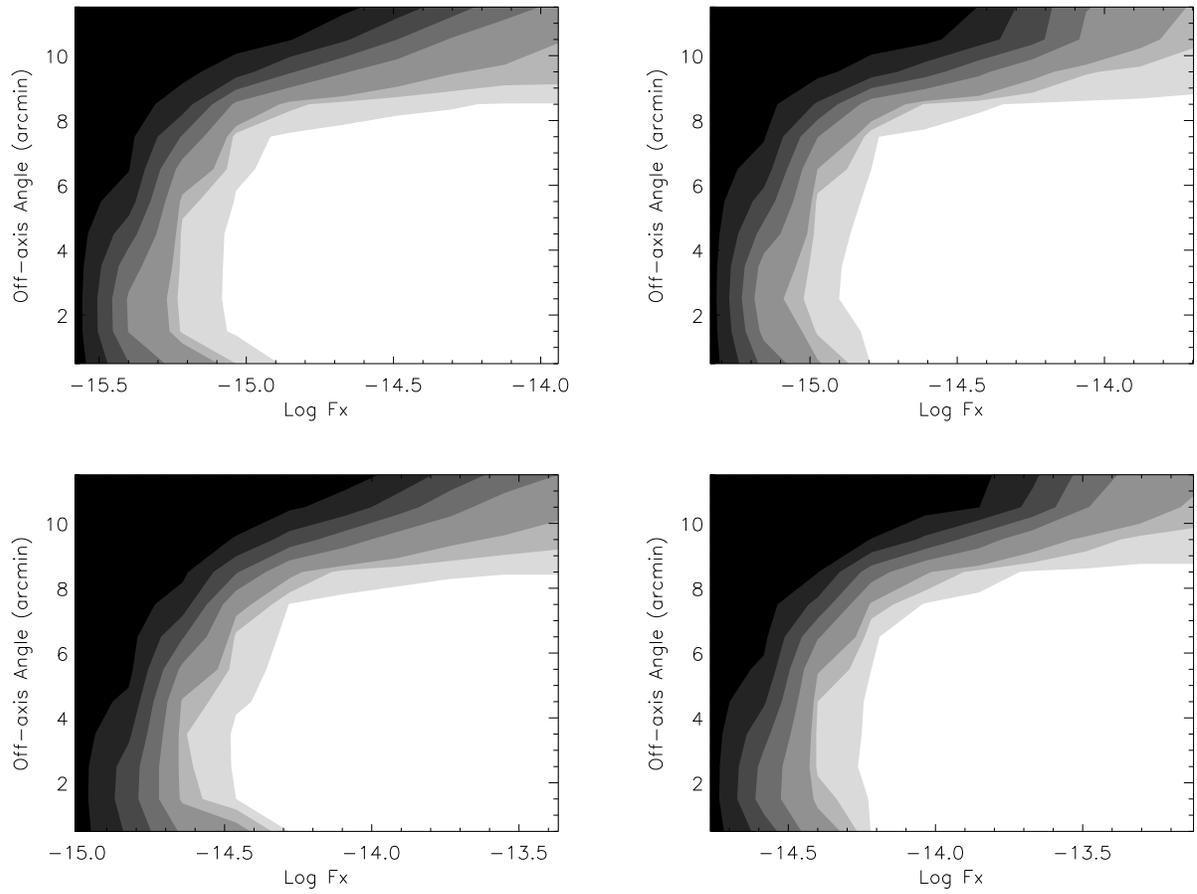}
\caption[fig3.ps] {The probability of source detection as a function of 
off-axis angle and 2--8~keV fluxes. Contour levels are 0.1,0.3,0.5,0.7,0.9,
0.95,0.99. Upper(lower) panels: soft (hard) band; Left (right) panels: 
70~ks exposures and 40~ks exposures.
\label{det_prob}}
\end{figure}

The {\it Chandra} detection sensitivity is not uniform because of vignetting effects,
quantum efficiency changes across the field and the broadening of the
point spread functions. The consequence is that the 
sensitivity of source detection drops monotonically with off-axis angles. 
To quantify this we generate simulated observations of our 40 ks and 70 ks exposure in 
both soft and  hard bands. Using {\it wavdetect} ({Freeman} {et~al.} 2002) on these 
images we obtain 
an estimate of the detection probability function at different fluxes and 
off-axis angles (Figure~\ref{det_prob}).

\begin{figure}
\includegraphics[scale=.7,angle=90]{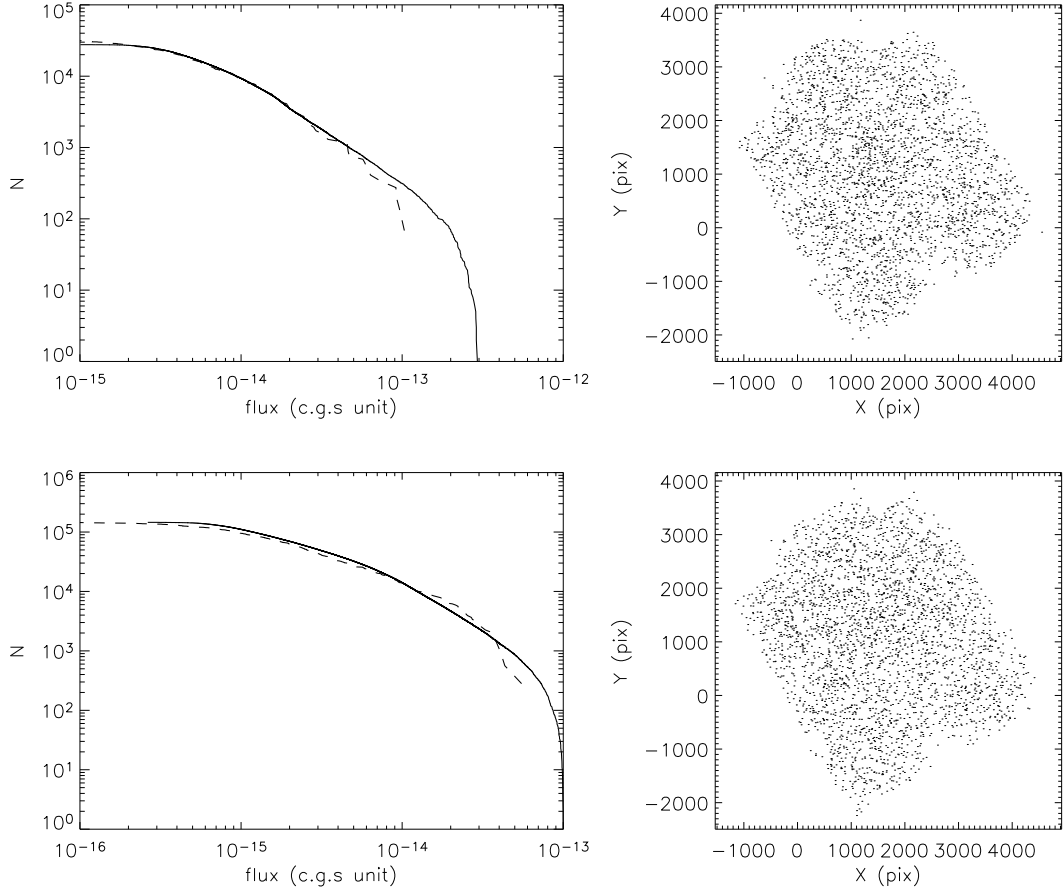}
\caption[fig4.ps]{The right panels shows the random sources after 
detections. The pixel size is 0.492\arcsec. The left panels show the cumulated 
counts of the simulated sources (solid line) and that of the observed sources 
(dashed line). Top panels: hard band; bottom panels: soft band.
\label{simulation}}
\end{figure}

With this probability, we can generate randomly distributed sources
with the X-ray selection effects to the first order. We use
this method instead of running detections on a large number of simulated
images because the detection program runs very slowly 
on these images. We generate source fluxes based on the best fit LogN-LogS 
from {Yang} {et~al.} (2004) and then ``detections'' are run on each of the images.
The resulting catalogs from all the nine simulated images are then merged in
the same way as for the real data. The resulting random source 
distribution and the resulting cumulative counts are shown in Figure~\ref{simulation}.

\begin{figure}
\includegraphics[scale=.7,angle=90]{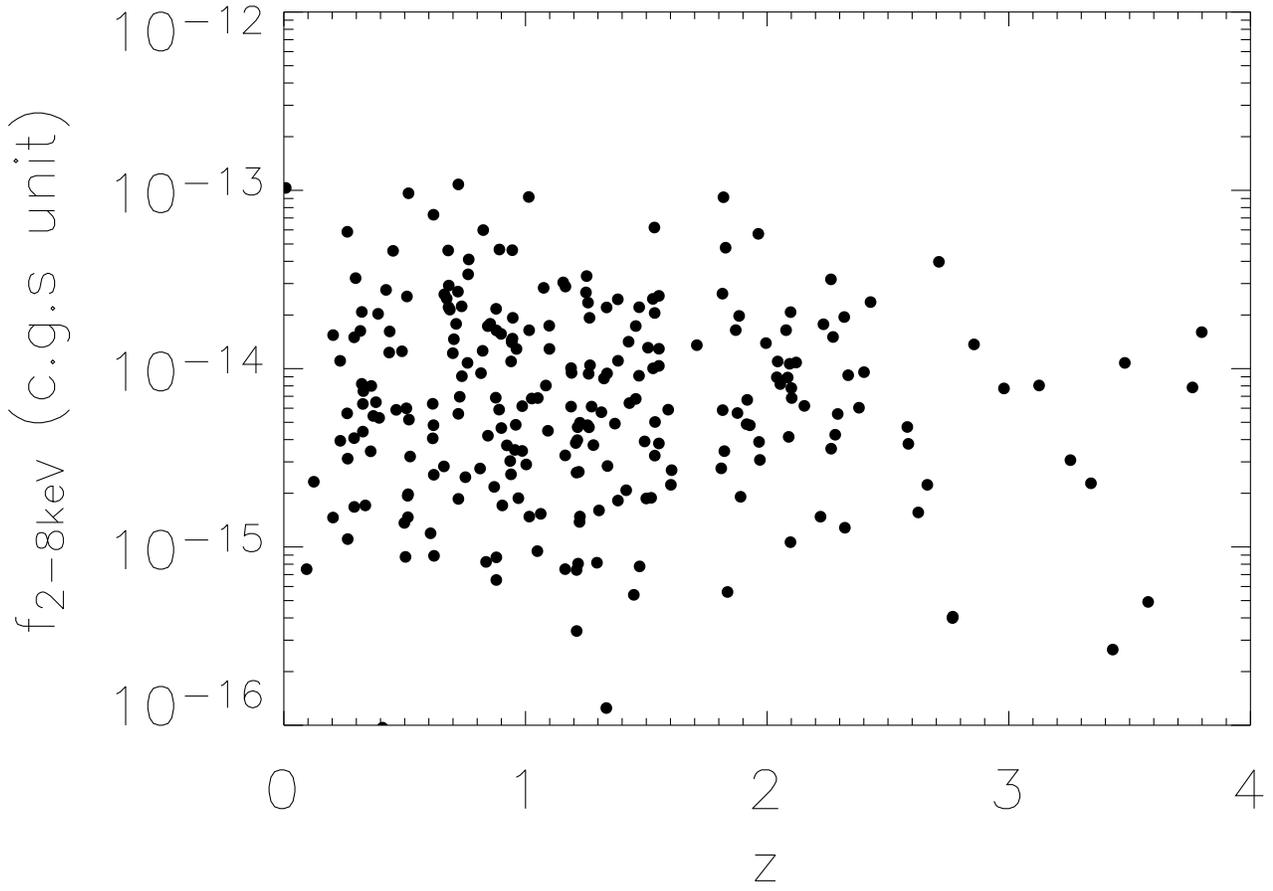}
\caption[z_flux.ps] {The 2--8~keV flux vs. redshift in CLASXS sample. 
There is no significant correlation between X-ray flux and redshift. 
\label{z_flux}}
\end{figure}

We next consider the optical selection effects. Since our spectroscopic
observation is close to complete for all sources with R$<24.5$, the sky
coverage is uniform and only a very small number of sources which are very 
close to each other could be missed. This might reduce the power at very small scales. 
The redshift distribution of the sources shows a very weak 
dependence on the X-ray flux (Figure~\ref{z_flux}), which is due largely to the
very broad luminosity function of AGNs.
We can thus ``scramble'' the observed redshifts and assign them to the simulated
sample without introducing a significant bias. The major selection effect in 
our optical observation is that the optical identifications are biased toward 
brighter sources. We select the simulated sources based on their X-ray flux 
using the best-fit curve in Figure~\ref{optical_id} as a probability function. 
The optical selection removes a large fraction of X-ray 
dim sources and therefore reduces the non-uniformity in the angular distribution 
caused by the X-ray selection effects. 
The redshift of the random sources were 
sampled from a Gaussian smoothed ($\sigma_{z} = 0.2$) redshift distribution 
from the observations. The purpose of the smoothing is to remove possible 
redshift clustering in the random sample but still preserve the effect of
the selection function. We tested different smoothing scales 
$\Delta z = 0.1-0.3$ and found the resulting correlation function 
effectively unchanged.  

\section{Results
\label{sec_results}}
\subsection{Redshift-space correlation function}
We calculate the redshift-space correlation function in the CLASXS sample 
for non-stellar sources with $0.1<z<3$  and 2--8~keV fluxes $> 5 \times 10^{-16}$\ergpcmsqps,
assuming constant clustering in comoving coordinates. The total number of 
sources in the sample is 233. The median redshift of the sample is 1.2. 
By comparing the number of detected pairs with separations $<20$~Mpc with that expected 
by simulation,  we found that on scales of 20~Mpc, the significance of clustering 
is $6.7\sigma$. 

We use the maximum likelihood method in searching for the best-fit parameters 
({Cash} 1979; {Popowski} {et~al.} 1998; {Mullis} {et~al.} 2004).  The method is 
preferable to the commonly used $\chi^{2}$ method because it is less 
affected by arbitrary binning. The method uses very small bins so that each 
bin contains only 1 or 0 DD pair. In this limit, the  probability associated 
with each bin is close to independent. The expected number of DD 
pairs in each bin is calculated using the DR, RR pairs using the mock catalog. 
The likelihood is defined as 
\begin{equation}
{\mathcal L} \equiv \prod_{i} \frac{e^{-\mu_{i}} \mu_{i}^{x_{i}}}{x_{i}!}
\end{equation}
where $\mu_{i}$ is the expected number of pairs in each bin and $x_{i}$
is the observed number of pairs. The likelihood ratio defined as
\begin{equation}
S \equiv -2 \ln ({\L} /{\L}_{0})
\end{equation}
where $\L_{0}$ is the maximum likelihood. The resulting $S$ 
approaches the usual $\chi^{2}$ distribution. Since the maximum-likelihood does 
not provide a measure of the ``goodness-of-fit'', we quote the $\chi^2$ derived from the 
binned correlation function (as shown in the figures) and the best-fit
parameters from maximum-likelihood estimates. 
 
\begin{figure}
\includegraphics[scale=.55,angle=90]{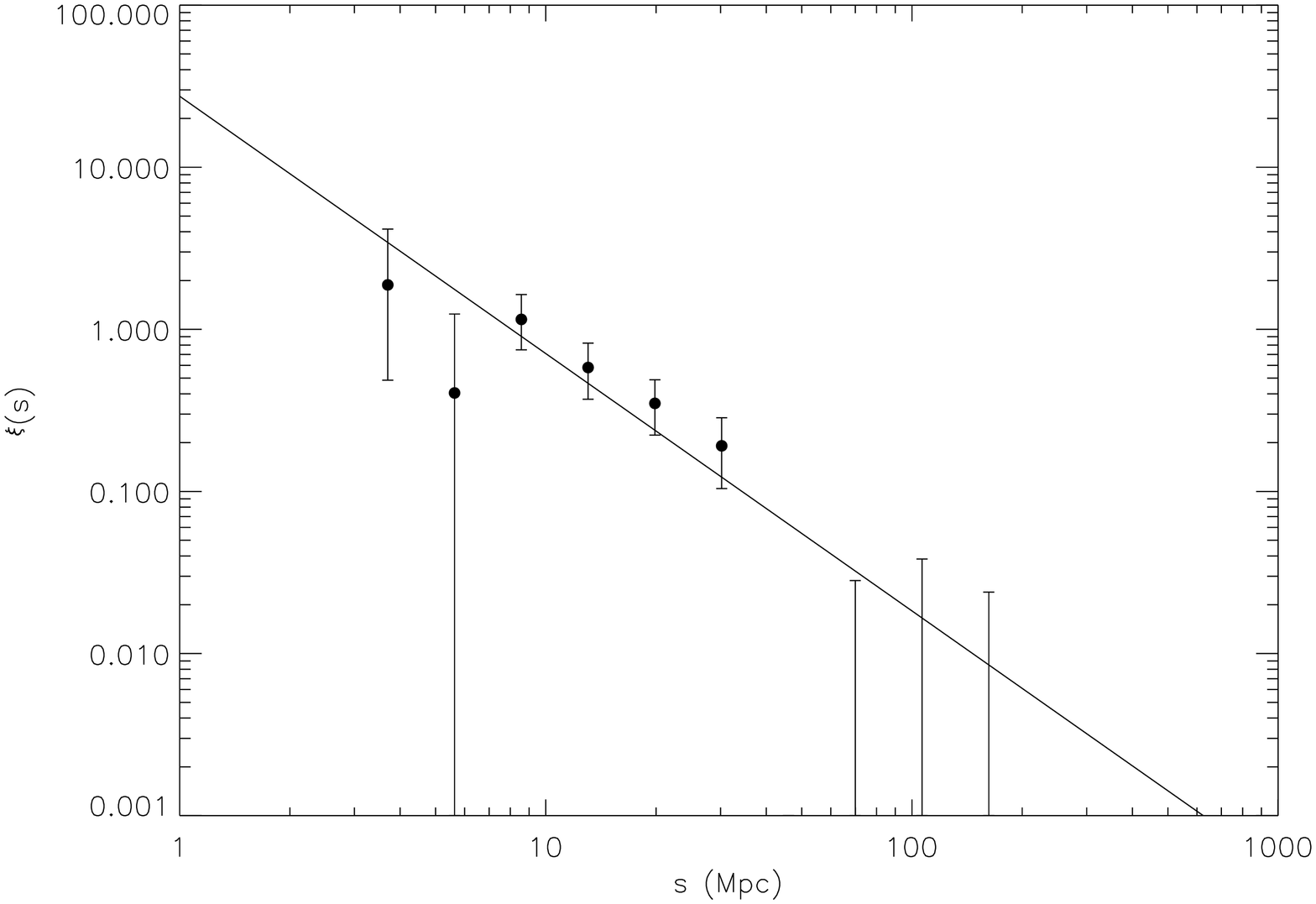}
\includegraphics[scale=.55,angle=90]{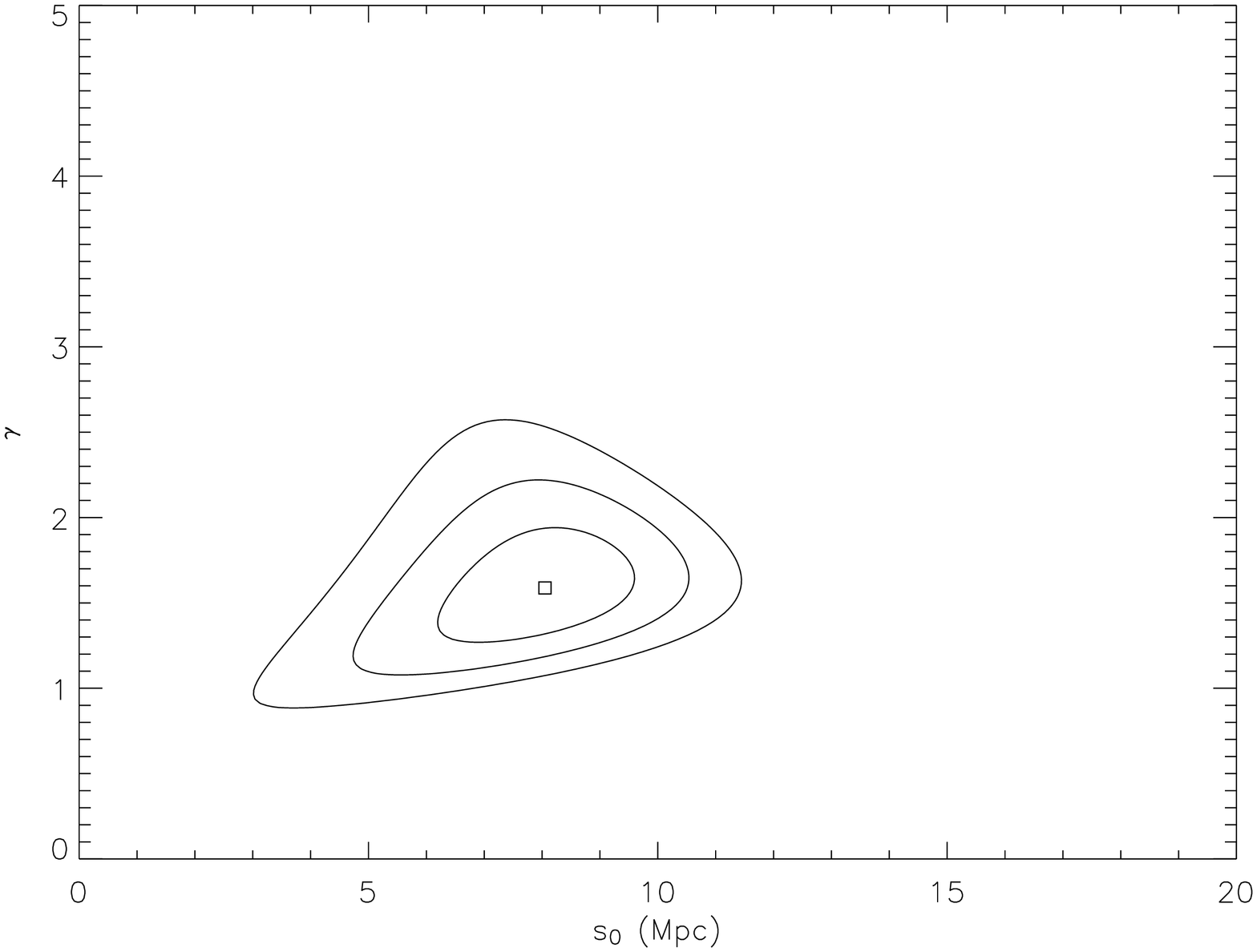}
\caption[acf_clasxs_3to200.ps,cont_clasxs_3to200.ps] {(a). The redshift-space
correlation function for CLASXS field with 3~Mpc$<s<$200~Mpc. 
(b). Maximum-likelihood contour for the single power-law fit. 
Contour levels are $\Delta S = 2.3, 6.17, 11.8$, corresponding to 
$1\sigma$, $2\sigma$ and $3\sigma$ confident levels for two parameter fits.   
\label{clasxs_cor}}
\end{figure}

We fit the correlation functions over three separation ranges. 
In Figure~\ref{clasxs_cor} we show the correlation function and the best-fit with 
3~Mpc$<s<$200~Mpc.  The best-fit parameters for all three separation ranges 
are listed in Table~\ref{z_space_cor}. 
It is noticeable that the rather large $\chi^2$ seems to suggest that the single
power-law model may not be a proper description of the data. 

\begin{figure}
\includegraphics[scale=.6,angle=90]{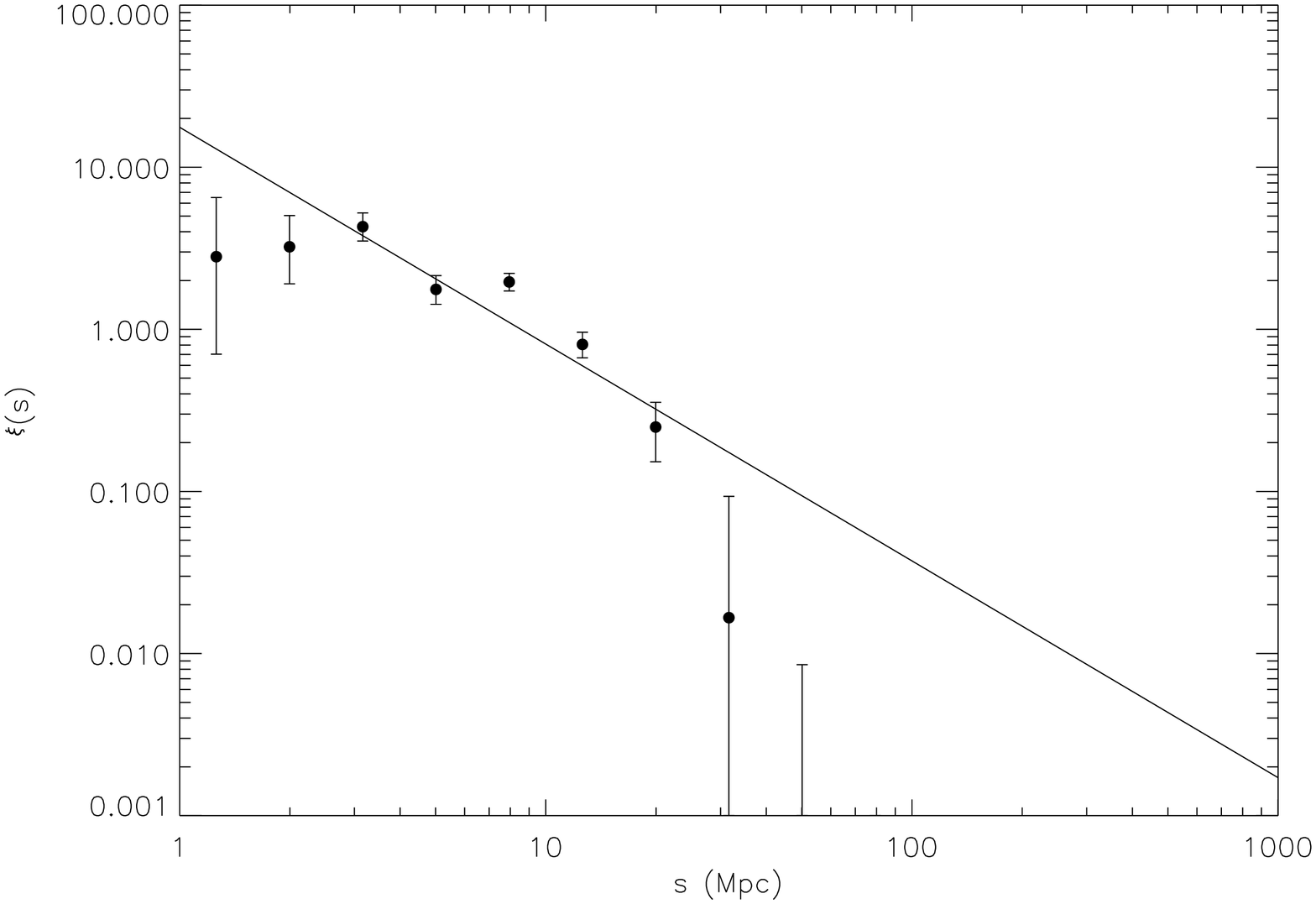}
\includegraphics[scale=.6,angle=90]{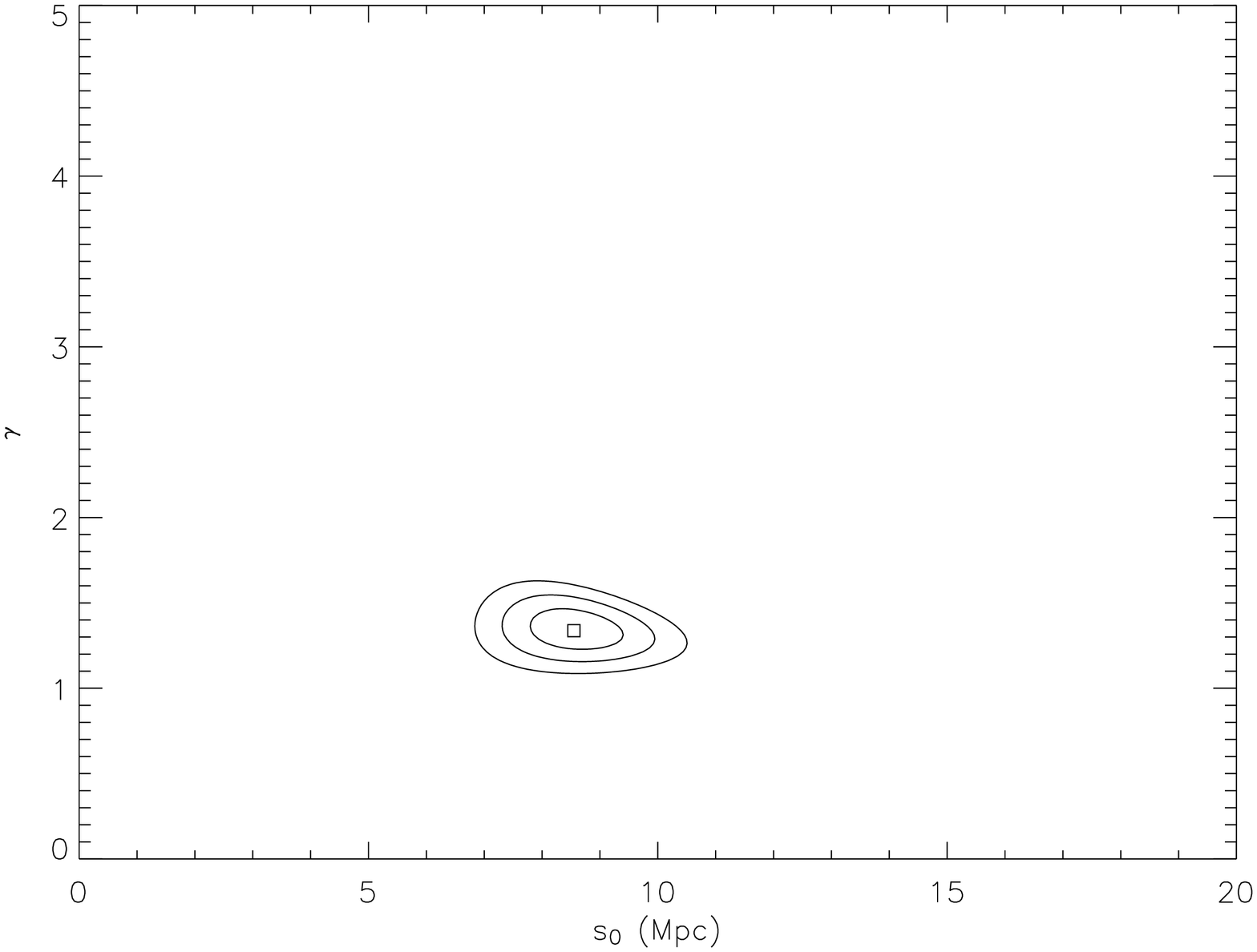}
\caption[acf_cdfn_1to100.ps,cont_cdfn_1to100.ps] {The same as Figure 
\ref{clasxs_cor} for the CDFN except that the correlation function is calculated
for separations 1~Mpc$<s<$100~Mpc.
\label{cdfn_cor}}
\end{figure}

For comparison, we also computed the 
correlation function of the X-ray sources in CDFN 
in the same redshift interval. We created a mock catalog 50 times 
larger than the observation. The positions and redshifts of the random 
sources are generated by randomizing the observed positions and redshifts. 
A large Poisson noise was added to avoid artificial clustering in the mock catalog.
Such randomization is justified because the clustering signal in a small 
field like the CDFN mainly comes from clustering along the line-of-sight direction. 
The randomized sky coordinates are filtered using an image mask to take 
into account the edge effects. We include all the non-stellar sources in the 
same redshift interval as we use for CLASXS, which results in 252 sources
in the sample. The best-fit parameters 
for CDFN field over three scale ranges are also shown in Table~\ref{z_space_cor}.
The correlation function over  1~Mpc$<s<$100~Mpc is shown in Figure~\ref{cdfn_cor}. 

There is a good agreement of the correlation lengths obtained
in the two fields.  There seems to be a systematic flattening 
of the slope at small separations ($s \sim 10$~Mpc) in both samples.
When the correlation functions are fitted at small and large separations
independently, the resulting $\chi^2$s are systematically smaller.   

\subsection{Projected correlation function
\label{sec_projected}}
\begin{figure}
\includegraphics[scale=.7,angle=90]{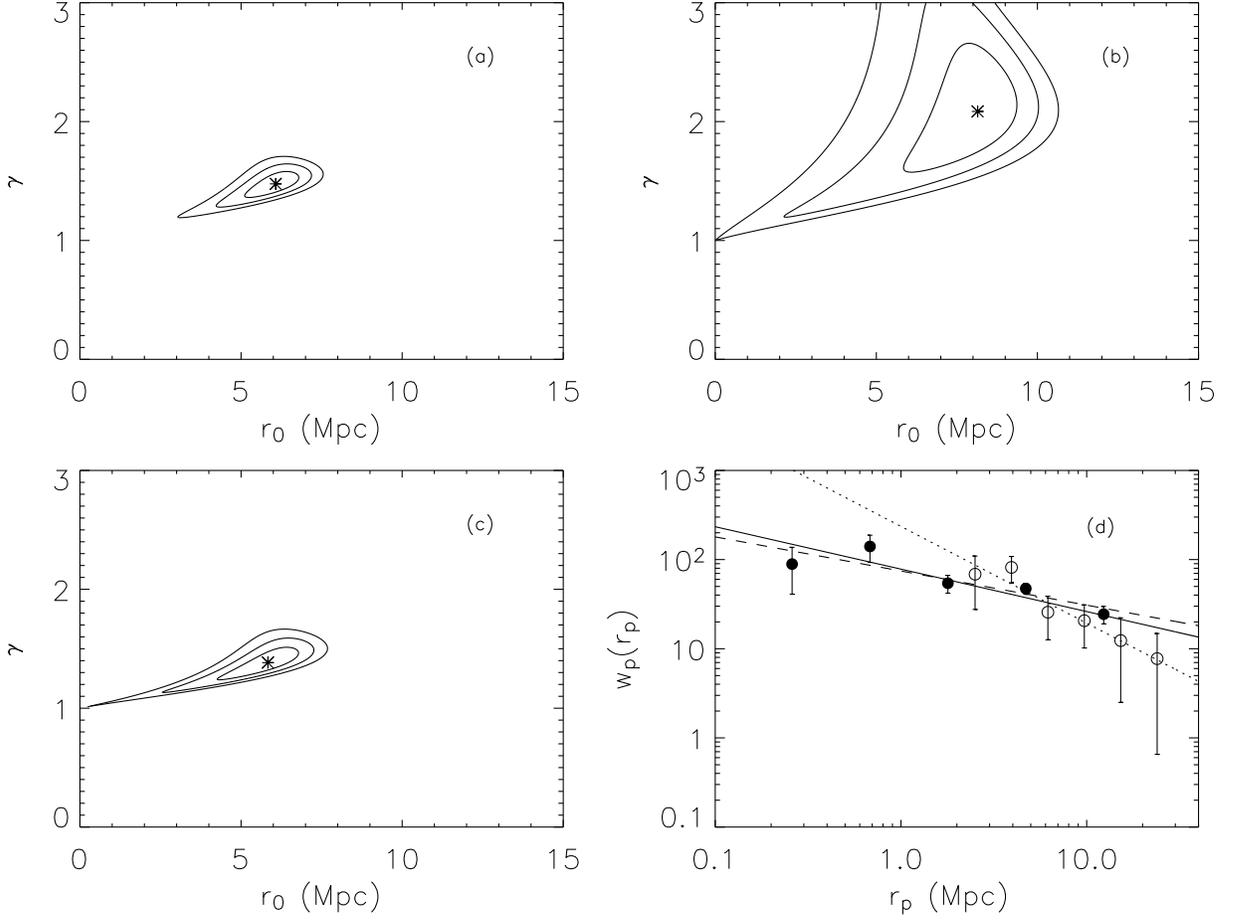}
\caption[Projected_all.ps] {The projected correlation function
for the CLASXS, CDFN fields and the combined sample. (a)-(c) show the best-fit parameters as well as 
the $\chi^2$ contours for the CLASXS+CDFN sample, the CLASXS sample, and the CDFN sample, respectively. 
Contour levels are for $1\sigma$, $2\sigma$, and $3\sigma$ confident levels; 
(d) The projected correlation function for CLASXS (open circles) and
CDFN (black dots) fields. Lines are the best-fit shown in (a)-(c).
Solid line: CLASXS+CDFN; Dotted line: CLASXS; Dashed line: CDFN
\label{Projected}}
\end{figure}

The projected correlation function is computed using the methods described
in \S~\ref{subsec_corr_func}. To test the method, we first compute the projected correlation function
for the CDFN and compare the results with that published in {Gilli} {et~al.} (2005). We selected 
the same redshift interval for the CLASXS field.  A two dimensional correlation function is 
calculated on a $5 \times 10$ grid on the ($r_p$,$\pi$) plane. The 5 intervals 
along $r_p$ axis covers 0.16--20~Mpc. We integrate the resulting two 
dimensional correlation function along the line-of-sight
to a $\pi_{max} = 20$~Mpc. Our projected correlation function for CDFN is shown 
Figure~\ref{Projected}, and it agrees perfectly with that reported in {Gilli} {et~al.} (2005) 
for $z = 0-4$, validating the techniques used in this paper.    

We next compute the projected correlation function for the CLASXS field.
The correlation function is calculated on scales of $r_p = 1 - 30$~Mpc. The 2-D correlation
function is integrated to $\pi_{max} = 30$~Mpc. The result is also shown in Figure~\ref{Projected}. 
The correlation functions of the CDFN and CLASXS fields agree in general 
at $r_p \sim 10$~Mpc, where both surveys have very good S/N. The slope, however, appears 
to be flatter in the CDFN field. 

We perform a $\chi^2$ fit to the correlation functions using Equation~\ref{wp_n_real}.  
The best-fit parameters for the CDFN are $r_0 = 5.8^{+1.0}_{-1.5}$~Mpc, $\gamma = 1.38^{+0.12}_{-0.14}$,
and the reduced $\chi^2/dof = 2.5/3$.  This is in good agreement with the result from Gilli et al. 
(2005, $r_0 = 5.7$~Mpc, $\gamma = 1.42$). The quoted errors in that paper is smaller than we obtained, 
but since we adopt a bootstrap error instead of Poisson error in this analysis, the difference is expected.
The best-fit parameters for the CLASXS field are $r_0 = 8.1^{+1.2}_{-2.2}$~Mpc, $\gamma = 2.1^{+0.5}_{-0.5}$, 
and the reduced $\chi^2/dof = 1.6/4$. The correlation length appears to be higher than that of the CDFN, but 
agrees within the errors. The slope also seems steeper than that of the CDFN and agrees better with the slope 
of the redshift-space correlation function at $r_p> 10$~Mpc. Since the CLASXS sample does not cover
separations $< 10$~Mpc very well, it is hard to see a slope change in this sample alone. 
Since the CDFN and CLASXS connect very well at separations where both surveys are sensitive, 
we try to model the combined data points with a single power-law. This yields 
$r_0 = 6.1^{+0.4}_{-1.0}$~Mpc, $\gamma_0=1.47^{+0.07}_{-0.10}$, and  $\chi^2/dof = 10.7/9$.  
The reduced $\chi^2$ is much worse than the two samples fitted separately, but does not 
reject the hypothesis of a single power-law fit. This seems to suggest that the slope 
of the correlation function flattens at small separations. Such a trend need to be looked more 
closely with future large AGN surveys with better signal-to-noise.  

\subsection{Redshift distortion
\label{sec_z_distortion}}
Redshift distortion affects the correlation function (power-spectrum) by 
increasing the redshift-space correlation amplitude and changing the shape
of the 2-D redshift-space correlation function at small scales (such as the 
well known ``finger-of-God'' effect, e.g. {Hamilton} 1992).
Since our data is too noisy at small separations to detect the effect, we only discuss 
the effect of the amplitude boosting of correlation function in redshift-space.  
{Kaiser} (1987) showed that to the first order, 
\begin{equation}
\xi(s) = \xi(r)(1+\frac{2}{3}\beta+\frac{1}{5}\beta^2),
\label{distortion}
\end{equation}
where $\beta \approx \Omega_M(z)^{0.6}/b(z)$ and $b(z)$ is bias. 
In principle, the redshift-space distortion can be estimated
by comparing $\xi(s)$ and $\xi(r)$. To quantify the effect, we use the 
correlation function estimates at scales where both projected and redshift-space 
correlation functions are well determined. For the CDFN, we chose the correlation
function estimates at 10~Mpc and find $\xi(s=10~Mpc)/\xi(r=10~Mpc) = 1.75 \pm 0.55$, 
if the best-fit of $\xi(s)$ on 1-100~Mpc is used. The choice of this 
scale is justified given that the slope possibly changes below and above 10~Mpc,
as seen in the projected correlation function. Since the slope of the redshift- 
and real-space correlation function is very similar in the CDFN, the ratio is 
almost constant. For the CLASXS field, we chose to estimate the ratio at 20~Mpc,
where the S/N is the best.  
We find  $\xi(s=20~Mpc)/\xi(r=20~Mpc) = 1.73 \pm 0.42$ by using the best-fit 
on 1--100~Mpc for $\xi(s)$. The ratio changes slowly with the scales probed, 
but is within the errors. We find a general agreement between 
CLASXS and CDFN. To avoid the arbitrary choice of scales, and to make  
the best use of the data, we combine the two samples to study the redshift distortion 
effect on $\xi(r_p,\pi)$. Since the projected correlation function of CDFN and and CLASXS   
agree in general, we are encouraged to assume that the the two samples, even 
with the vast difference in flux limits, generally trace the large scale structure
in the same way.

\begin{figure}
\includegraphics[scale=.7,angle=90]{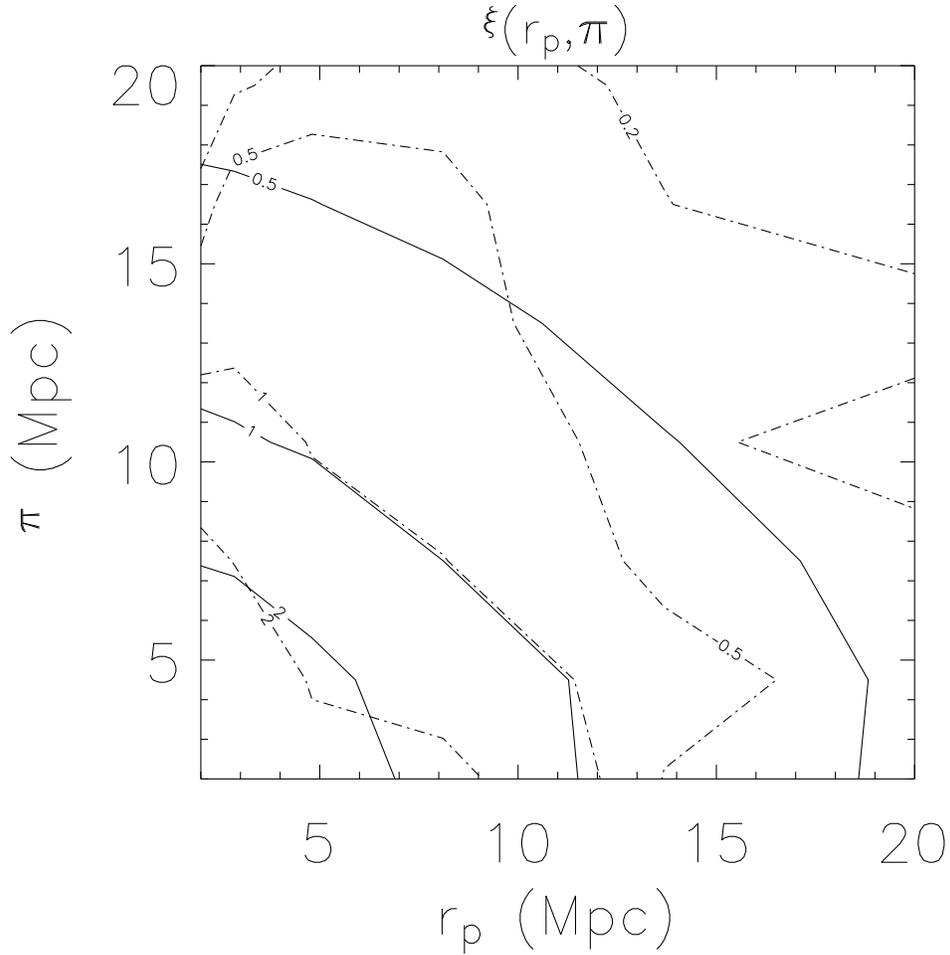}
\caption[z_distortion.ps] {Two dimensional redshift-space correlation 
function $\xi(r_p,\pi)$ of the combined CLASXS and CDFN sample (dashed-dotted
contour). Solid line shows the best-fit model. Both the data and model
correlation functions are smoothed using a $2 \times 2$ boxcar to reduce
the noise for visualization only. 
\label{z_distortion}}
\end{figure}

In Figure~\ref{z_distortion}. we show the combined  $\xi(r_p,\pi)$. The contours show 
no significant signature of nonlinear redshift distortion, such as the ``finger-of-god''. 
We fit $\xi(r_p,\pi)$ with Equation~\ref{distortion}, assuming the 
best-fit parameters for the real-space correlation function from the combined 
sample ($r_0 = 6.1$~Mpc, $\gamma_0=1.47$), and ignoring the higher order redshift 
distortions. We generate the 2-D correlation function at each grid point. 
By minimizing $\chi^2$ by changing $\beta$, we found the best-fit $\beta = 0.4 \pm 0.2$, 
corresponding to $\xi(s)/\xi(r) \sim 1.3$, which agrees with the estimates 
from individual fields above. 
By fixing $\Omega_M = 0.27$, we can estimate the bias factor of X-ray selected AGNs from $\beta$. 
The median redshift of the combined sample is 0.94, and 
$\Omega_M(z=0) = 0.27$ gives $\Omega_M(z=0.94) = 0.73$. This yields $b \approx 2.04 \pm 1.02$ using 
the relation $\beta \approx \Omega_M^{0.6}/b$.  
  
\subsection{X-ray color dependence}
We further test if there is any differences in clustering properties 
between the hard and soft spectra sources in the CLASXS sample. 
We use the hardness ratio, defined as 
$HR \equiv C_{2-8 keV}/C_{0.5-2 keV}$ (where $C$ is the count rate),
to quantify the spectral shape of the X-ray sources. Correlation functions
of soft ($HR < 0.7$) and hard ($HR \ge 0.7$) sources are calculated the same way 
as above. The fraction of broad-line AGNs is 56.4\% in the soft sample  
and 15.4\% in hard sample. The median redshifts are 1.25 and 0.94 for
soft and hard samples, respectively. We compute $\xi(s)$ for both soft and 
hard sources over scales of 3--200~Mpc.

Using a maximum-likelihood fit, we found $s_0 = 9.6^{+2.4}_{-3.4}$~Mpc,
$\gamma = 1.6^{+0.8}_{-0.6}$ for hard sources and  
$s_0 = 8.6^{+2.2}_{-2.0}$~Mpc, $\gamma = 1.6^{+0.6}_{-0.5}$ for soft
sources. We found no significant difference in clustering between the soft
and hard sources. This agrees with the results of Gilli et~al. (2005). It is noticeable
that the soft sources have a higher median redshift than the hard sources.
The interpretation of this result must include evolution effects. 
To avoid this complication, we restricted the redshift range to $z=0.1-1.5$.
The best-fit parameters are $s_0 = 9.5^{+3.1}_{-3.7}$~Mpc ($6.2^{+2.7}_{-4.6}$~Mpc)
and $\gamma = 1.7^{+0.9}_{-0.6}$ ($2.5^{+1.6}_{-0.9}$) for hard (soft) sources. 
The difference in clustering parameters between soft and hard sources are 
well within the measurement error. The same analysis on CDFN yields similar results. Thus there is 
no significant dependence of clustering on the X-ray color. 

\subsection{Luminosity dependence
\label{sec_L_dependence}}
The cold dark matter (CDM) model of hierarchical structure formation predicts that massive  
(and hence luminous) galaxies are formed in rare peaks, and therefore
should be  more strongly clustered. This is seen in normal galaxies 
(e.g. {Giavalisco} \& {Dickinson} 2001). Whether this relation can be extended to X-ray 
luminosity of AGNS is unknown. This is because the X-ray luminosity relates to the dark 
matter halo mass in a more complex and not well understood way. The X-ray luminosity is directly
linked to the accretion process, and the process is affected by factors such as
accretion rate, radiative efficiency, blackhole mass and the details of the accretion process. We have shown that at
least in broadline AGNs, where the blackhole mass can be inferred from the line-width
and nuclear luminosity, the Eddington ratio is close to constant over two decades
of 2--8~keV luminosity ({Barger} {et~al.} 2005). If this is the case for all X-ray 
selected AGNs, we should expect the AGN luminosity to be mainly determined by the 
blackhole mass, which in turn, should be closely related to the halo mass
({Ferrarese} 2002), even though the exact form of this relation is highly uncertain. 
However, optical quasar surveys such as 2dF found little evidence
of a correlation between clustering amplitude and ensemble luminosity (C05),
perhaps due to the small dynamical range in luminosity these surveys probe. 
The X-ray luminosity of sources in the CLASXS and CDFN 
cover a luminosity range of four orders of magnitudes, making it possible
to make such a test.  

\begin{figure}
\includegraphics[scale=.7,angle=90]{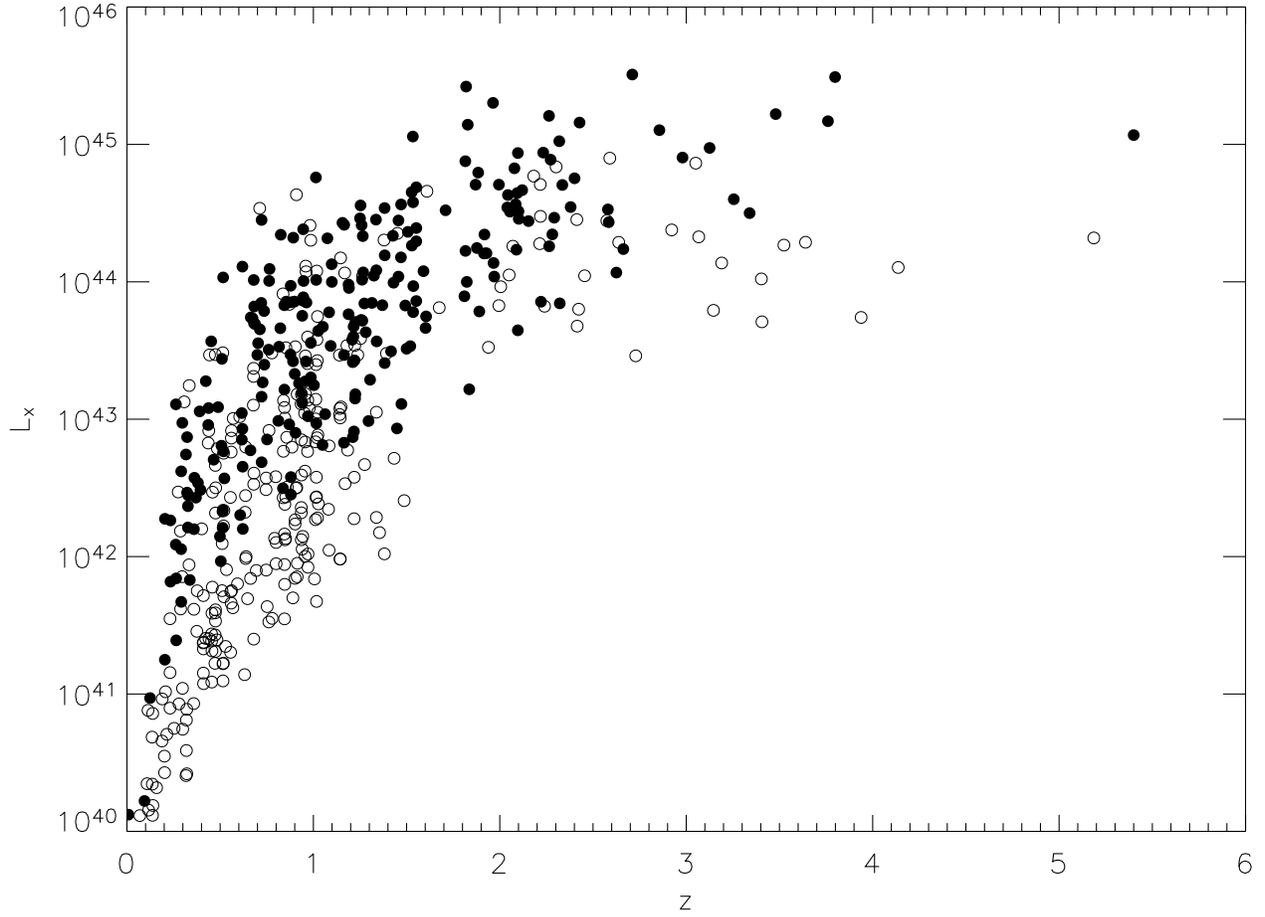}
\caption[Lum_z_LHNW_CDFN.ps]{The luminosity of X-ray sources vs. 
redshifts in the CLASXS (dots) and the CDFN (open circles) samples.
\label{Lum_dist}}
\end{figure}

The 2--8~keV rest frame luminosity $L_x$ is calculated from the hard band fluxes, with a K-correction
made assuming a power-law spectra with photon index $\Gamma = 1.8$. This yields
\begin{equation}
L_{x} = L_0 (1+z)^{0.2},
\label{k-cor}
\end{equation}
where $L_0$ is the luminosity in observer's 2-8~keV band.   
In Figure~\ref{Lum_dist} we show $L_x$ vs. redshift for both CLASXS and CDFN. 
For a better comparison
of the correlation amplitude, we adopt the averaged correlation function within $20h^{-1}$~Mpc,
\begin{equation}
\bar{\xi}(20) = \frac{3}{20^3}\int_{0}^{20} ds \xi(s)s^2. 
\label{avcor}
\end{equation}
The quantity is chosen rather than $s_0$ because it measures the clustering 
(directly linked to the rms fluctuations) regardless of the
shape of the correlation function. On scales of 20~Mpc the clustering is
in the linear regime of density fluctuations.  
The error in $\bar{\xi}(20)$  is from the single parameter $1\sigma$ confidence
interval obtained by fixing the slope of the correlation function to the best-fit.     

We split the CLASXS (CDFN) sample into two subsamples at $L_{x} = 4.5 \times 10^{43}$~\ergps 
($3.2 \times 10^{42}$~\ergps) so that each subsample contain similar number of objects. 
In Table~\ref{tab_lum} we show the
maximum-likelihood fits as well as $\bar{\xi}(20)$s. It should be noted that 
the correlation amplitude is biased in redshift space. The dominant part of this 
bias is characterized in Equation~\ref{distortion}. Comparing with other observations 
(e.g. {da {\^ A}ngela} {et~al.} 2005), $\beta$ is likely a weak function of redshift in the redshift 
range probed by our sample, 
with $\beta \sim 0.4$ (\S \ref{sec_z_distortion}), this translates to $\xi(s)/\xi(r) \sim 1.3$. 
We correct the $\bar{\xi}(20)$'s for this bias by
dividing them by 1.3. The correlation amplitude for the more luminous sources appears 
to be higher than that of the less luminous sources, which qualitatively agrees with 
expectations that X-ray luminosity reflects the dark matter halo mass.
The correlation amplitude for the more luminous subsamples are $2.3\sigma$ and 
$5.7\sigma$ higher than that of the less bright subsample in the CLASXS 
and CDFN fields, respectively. However, since the more luminous subsamples  
also are preferentially found at higher redshifts, the evolution in $\xi(s)$ 
should be taken into account. 

\begin{figure}
\includegraphics[scale=.7,angle=90]{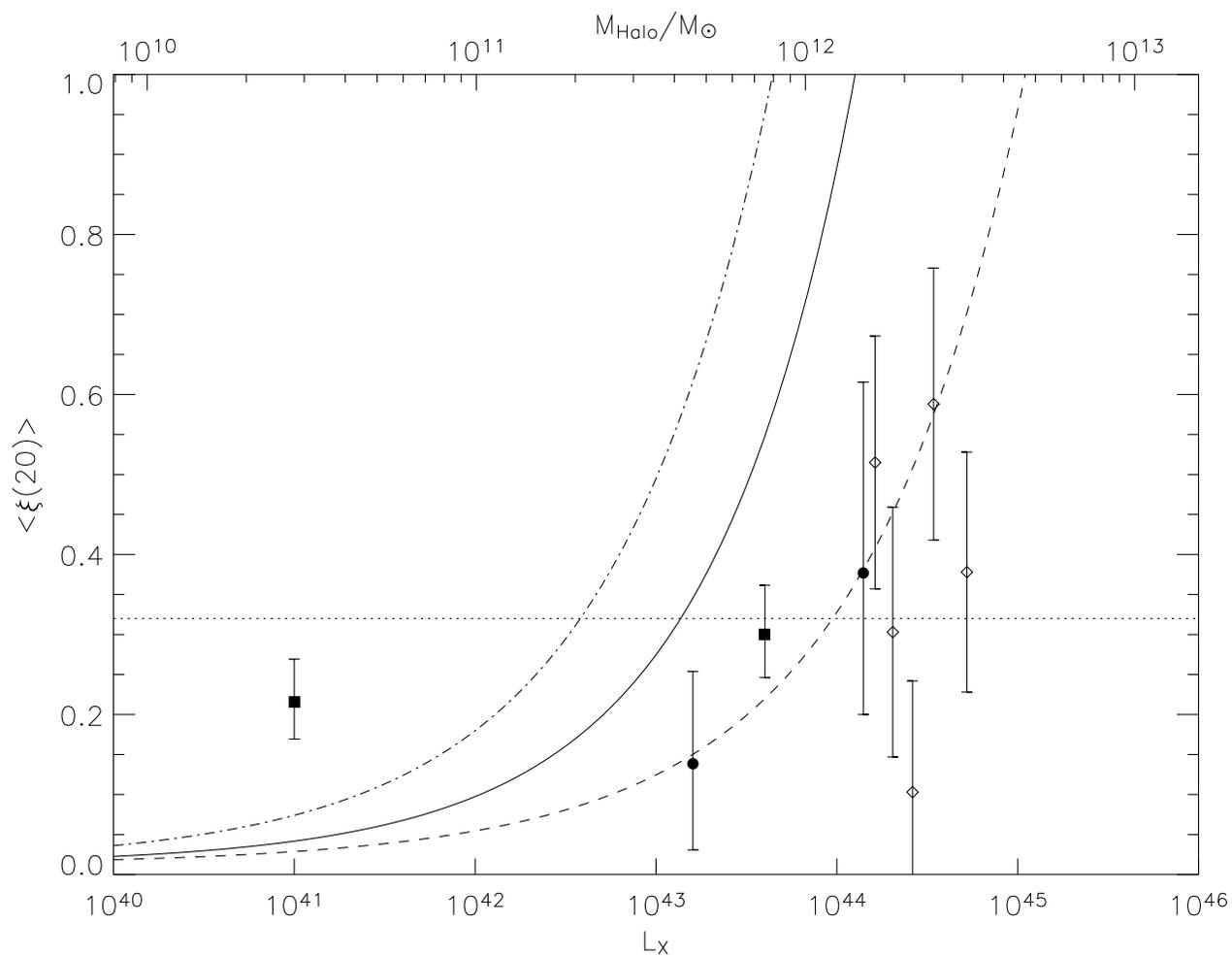}
\caption[lum_depend.ps]{Luminosity dependence of clustering
of AGNs. Black dots: CLASXS samples; Filled boxes: CDFN samples;
Diamonds: 2dF sample (Croom et al. 2004). Lines are the models
for different halo profile from Farrarese (2002). Solid line: 
NWF profile ($\kappa = 0.1$, $\lambda = 1.65$); 
Dashed line: weak lensing determined halo profile (Seljak, 2002;
$\kappa = 0.67$, $\lambda = 1.82$);
Dash-dotted line: isothermal model ($\kappa = 0.027$, $\lambda = 1.82$)
\label{Lum_depend}}
\end{figure}

To reduce this complication, we 
restrict ourselves to sources within the redshift range of 0.3--1.5, where the evolution 
effect is relatively small (see also \S~\ref{sec_evolution}). In Figure~\ref{Lum_depend} we show $L_x$ vs. 
$\bar{\xi}(20)$ for both CLASXS and CDFN. By reducing the redshift range, 
the difference in correlation amplitude between the brighter and dimmer 
subsample is significantly reduced, in the CDFN sample, to merely 
$1\sigma$.  
For the CLASXS field, on the other hand, the correlation amplitude for 
both subsamples do not show significant change. For comparison, we also 
plot in Figure~\ref{Lum_depend} the correlation amplitude from the 2dF survey (C05).
The X-ray luminosities for the QSOs in the 2dF are obtained by dividing the 
bolometric luminosities by 35 ({Elvis} {et~al.} 1994).    
We perform Spearman's $\rho$ test for correlations between $\log~L_x$ and 
$ \bar{\xi}$. We found the correlation coefficient $\rho=0.8$ for X-ray samples,
or a corresponding null probability of 20\%, indicating a weak correlation between
the two quantities. If the 2dF samples are added,  we found $\rho = 0.1$,
and a null probability of 17\%.  This means that with the X-ray sample, we have
detected a weak correlation between clustering and luminosity. We will discuss this in \S~\ref{sec_xi_Lx}.

\section{Evolution of clustering
\label{sec_evolution}}
Measuring the correlation function over a wide redshift range
only makes sense if the correlation function is a weak function
of redshift. The best measurements of clustering of 2dF quasars 
at high redshift show that the correlation function indeed exhibits
only mild evolution (C05). In this section, we test
the evolution of clustering of X-ray selected AGNs and compare 
them with other survey results. 

 \subsection{Chandra Sample}
\begin{figure}
\includegraphics[scale=.55,angle=90]{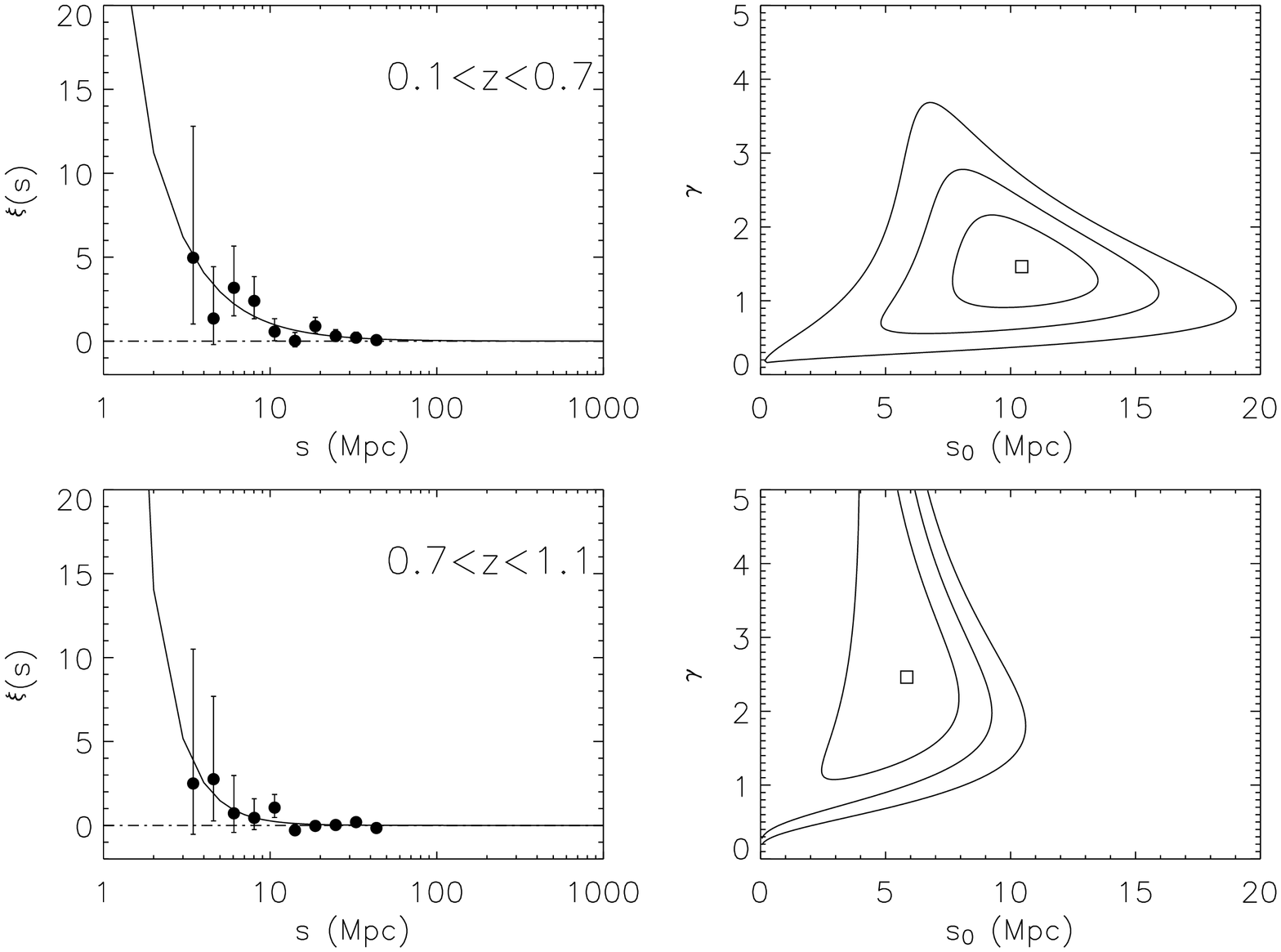}
\includegraphics[scale=.55,angle=90]{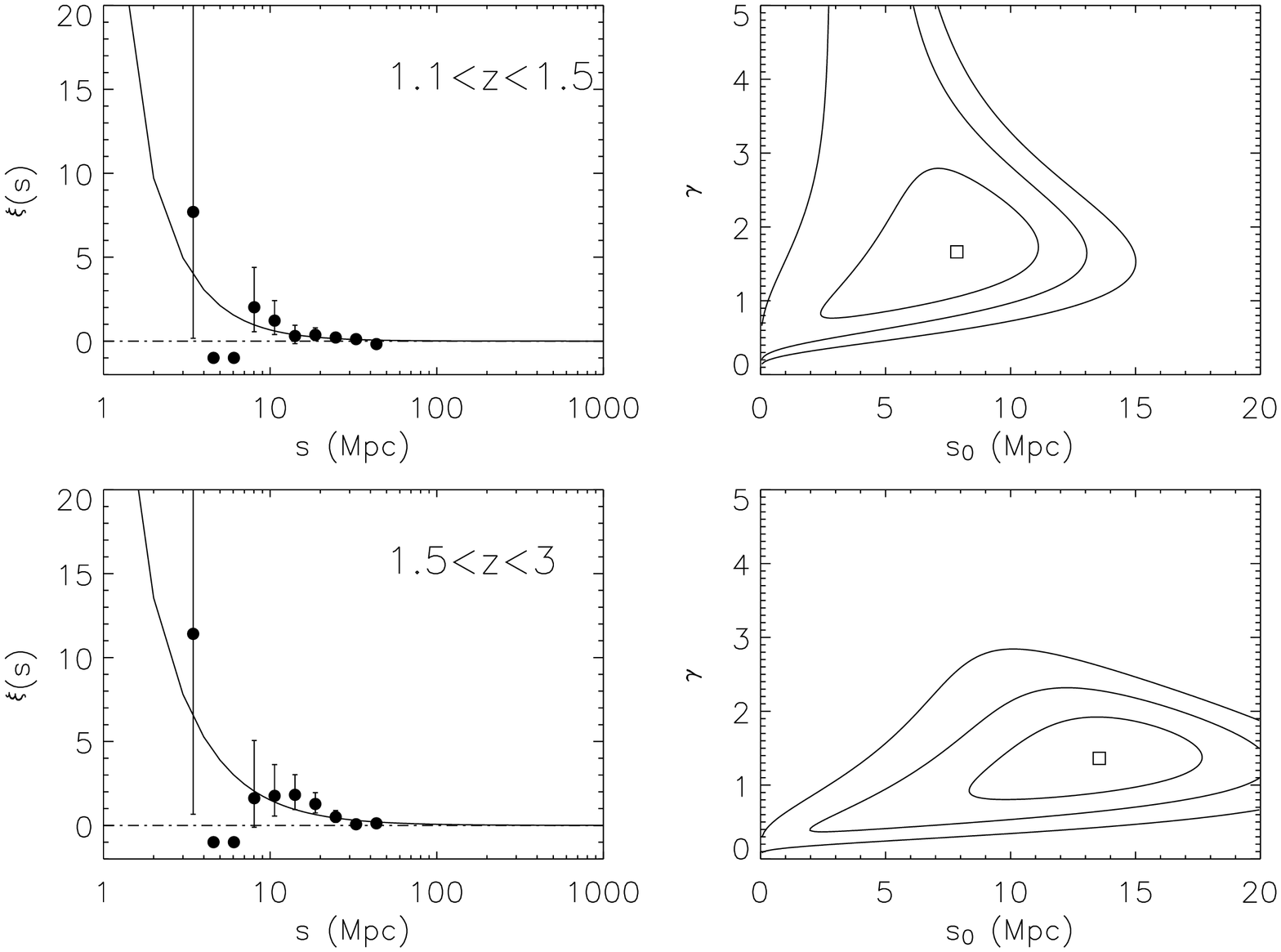}
\caption[z_depend_lhnw_1.ps,z_depend_lhnw_2.ps]{The Redshift-space
correlation function for the CLASXS field in four redshift bins. Left panels: 
The correlation functions and the power-law best-fits using maximum-likelihood
method. Right panels: the maximum-likelihood
contours for the corresponding correlation functions on the left. Contour
levels correspond to $1\sigma$, $2\sigma$ and $3\sigma$ confident levels.   
\label{z_dep_clasxs}}
\end{figure}

\begin{figure}
\includegraphics[scale=.6,angle=90]{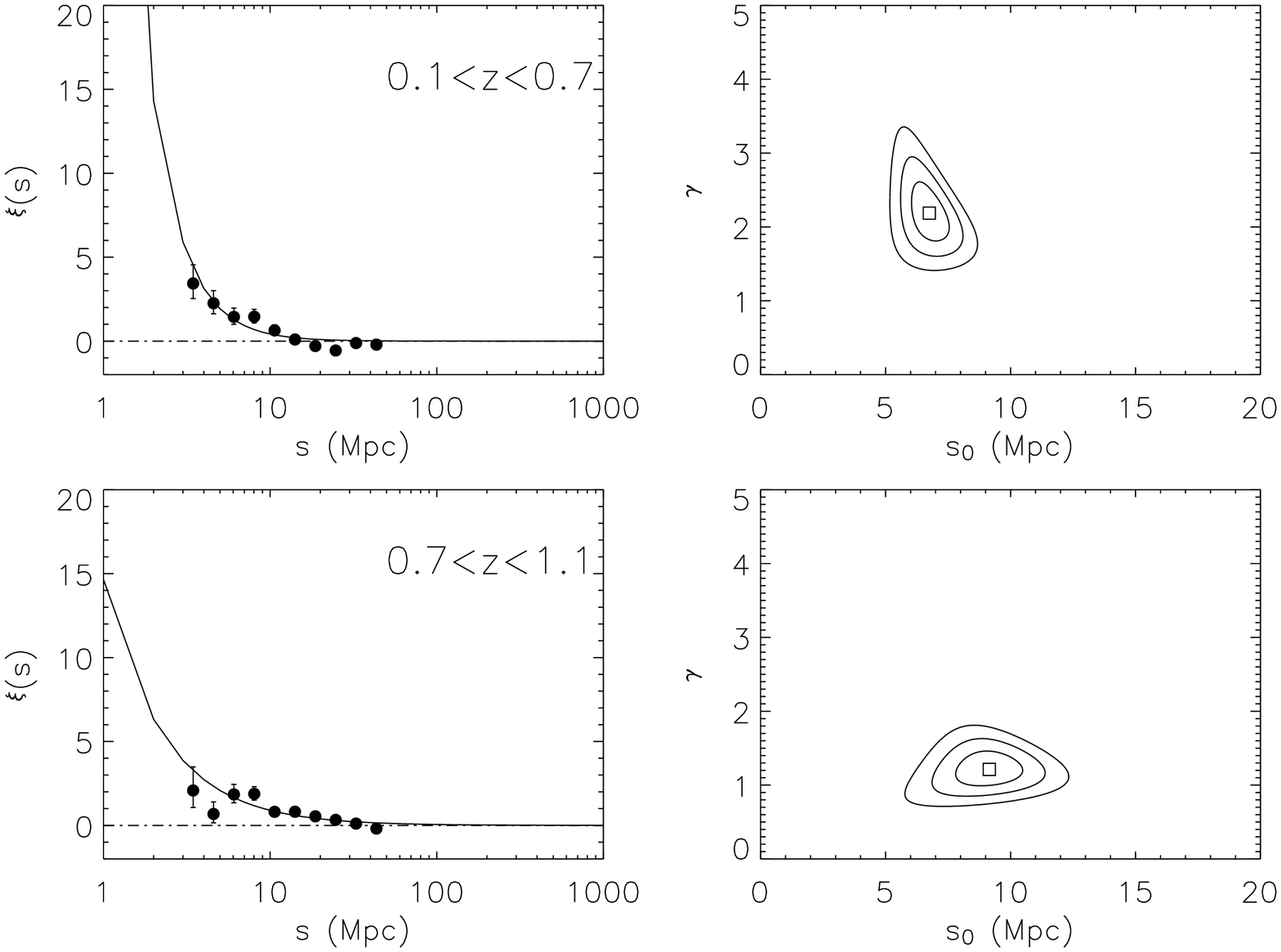}
\includegraphics[scale=.6,angle=90]{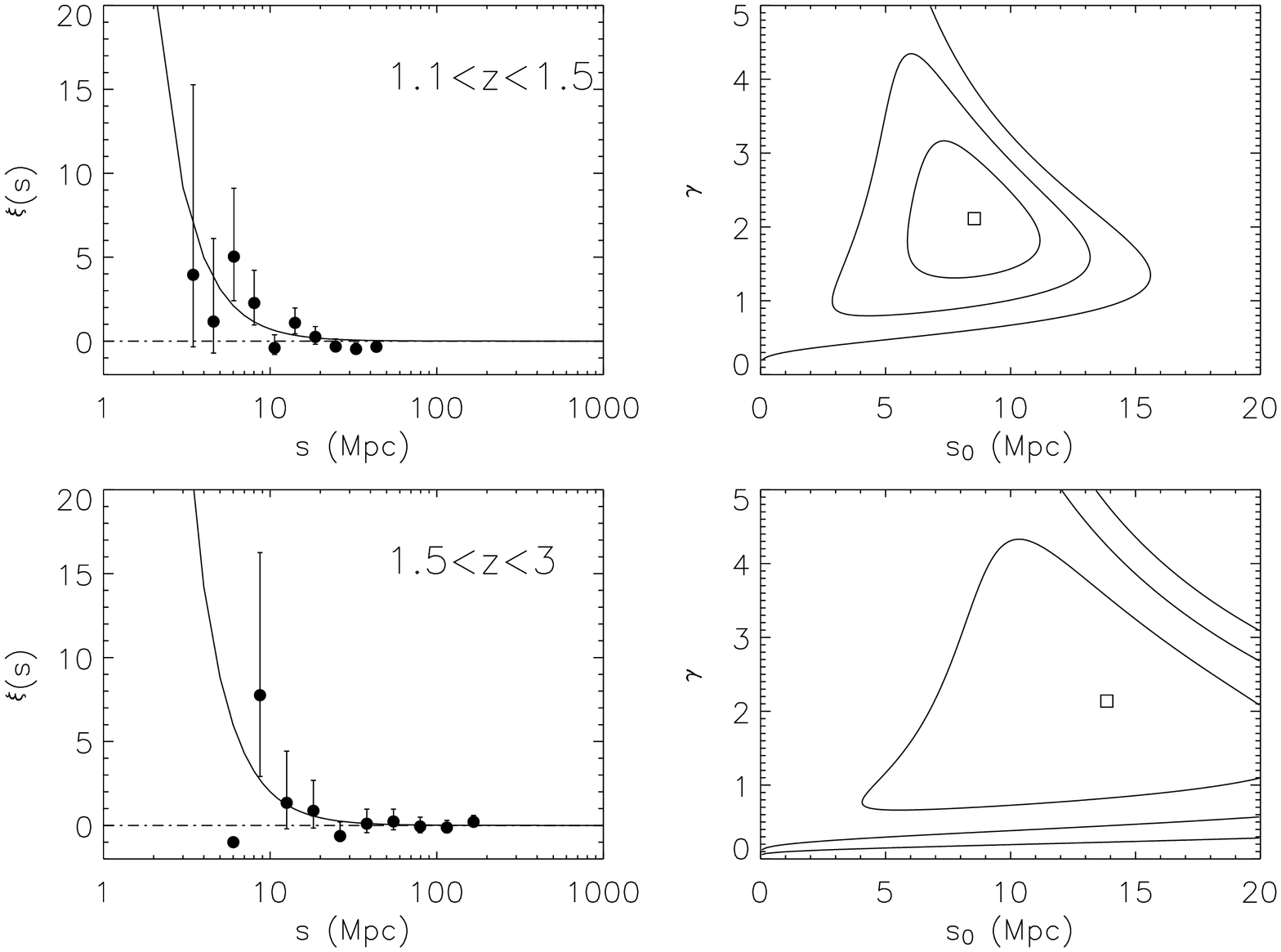}
\caption[z_depend_cdfn_1.ps,z_depend_cdfn_2.ps]{The Redshift-space
correlation function for the CDFN field in four redshift bins. (layout 
and contour levels are the same as in Figure \ref{z_dep_clasxs}).
\label{z_dep_cdfn}}
\end{figure}

\begin{figure}
\includegraphics[scale=.6,angle=90]{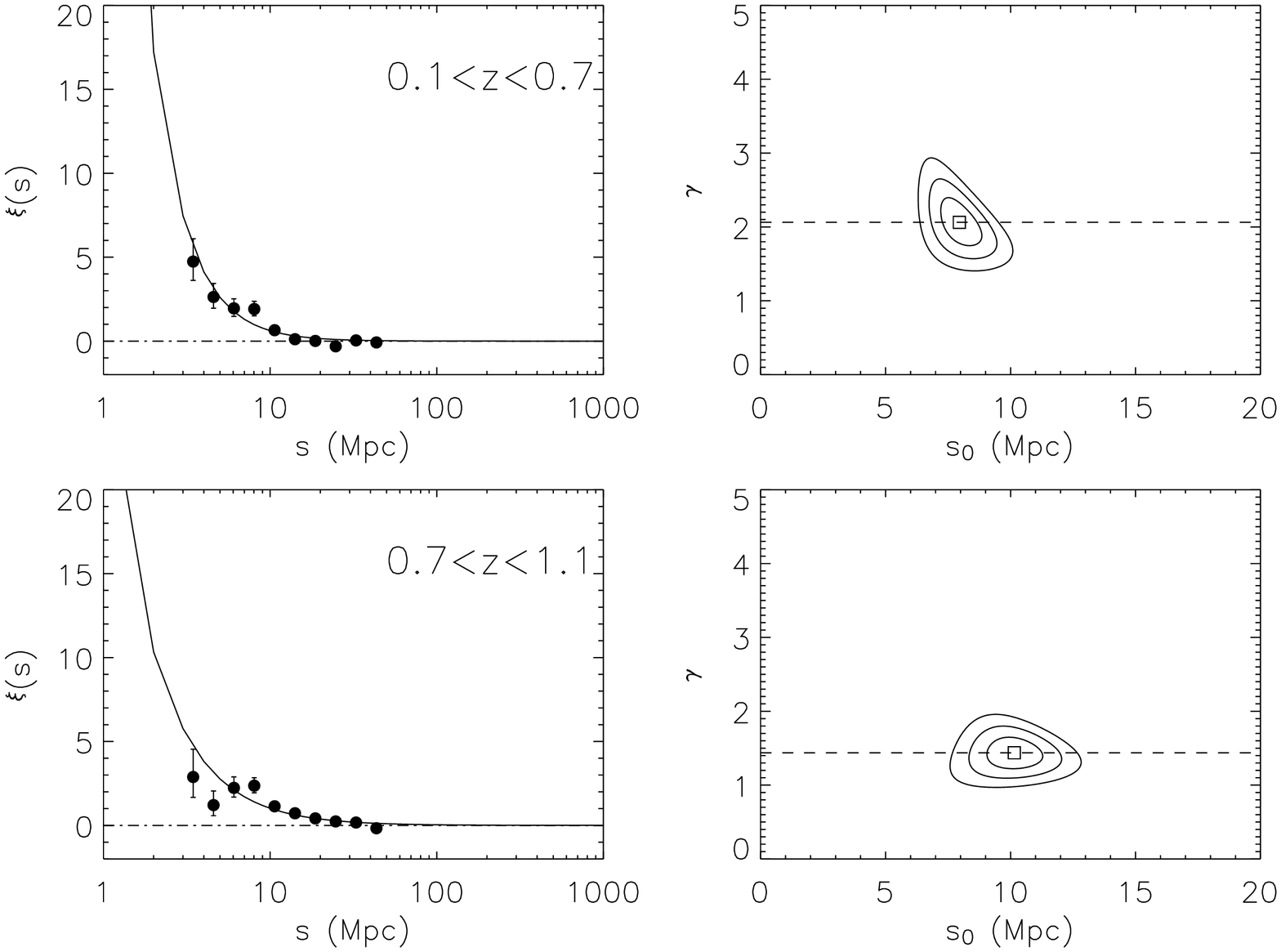}
\includegraphics[scale=.6,angle=90]{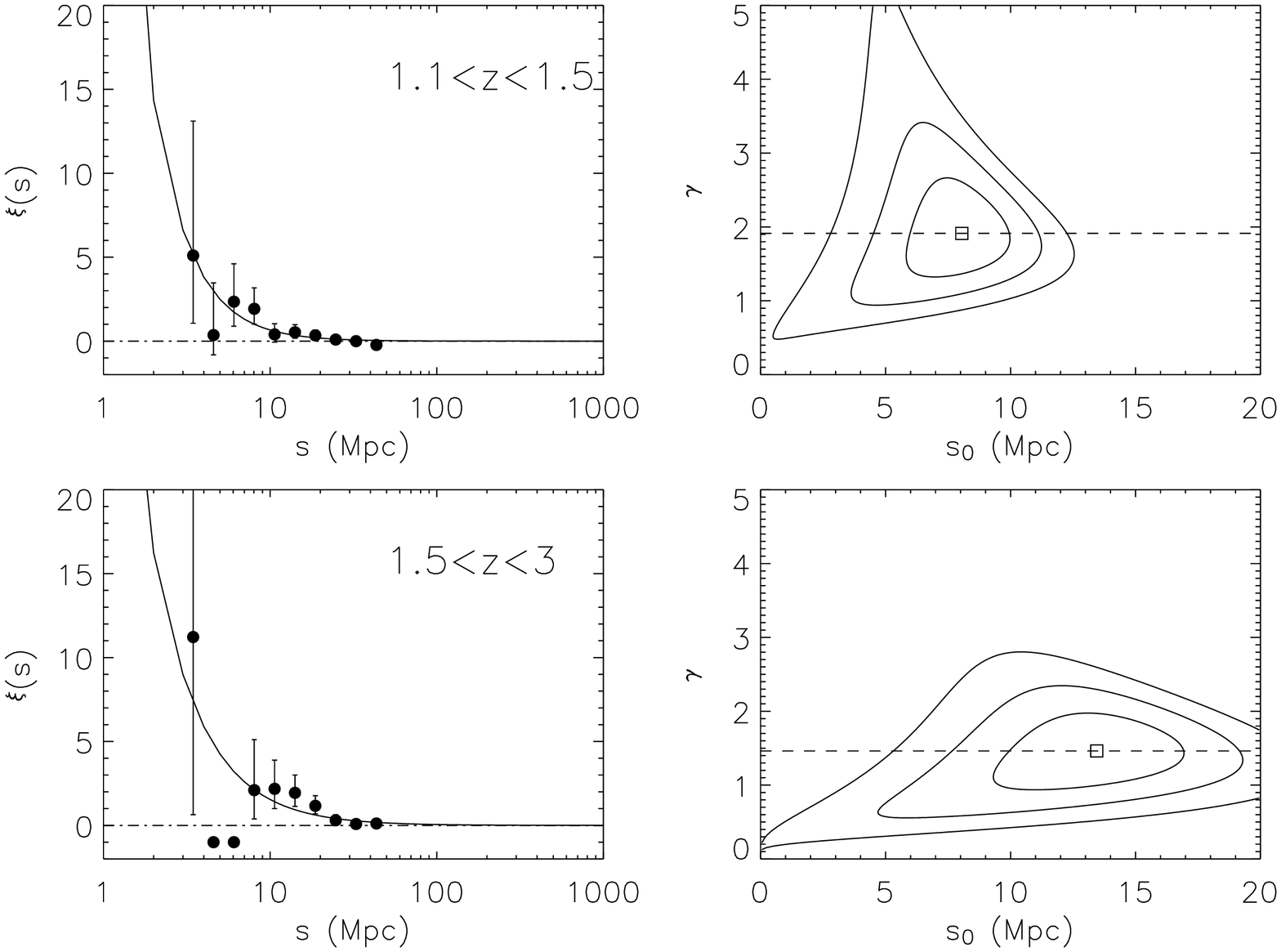}
\caption[z_depend_comb_1.ps,z_depend_comb_2.ps]{The Redshift-space
correlation function for the CLASXS+CDFN field in four redshift bins. (layout 
and contour levels are the same as in Figure \ref{z_dep_clasxs}).
\label{z_dep_comb}}
\end{figure}

\begin{figure}
\includegraphics[scale=.7,angle=90]{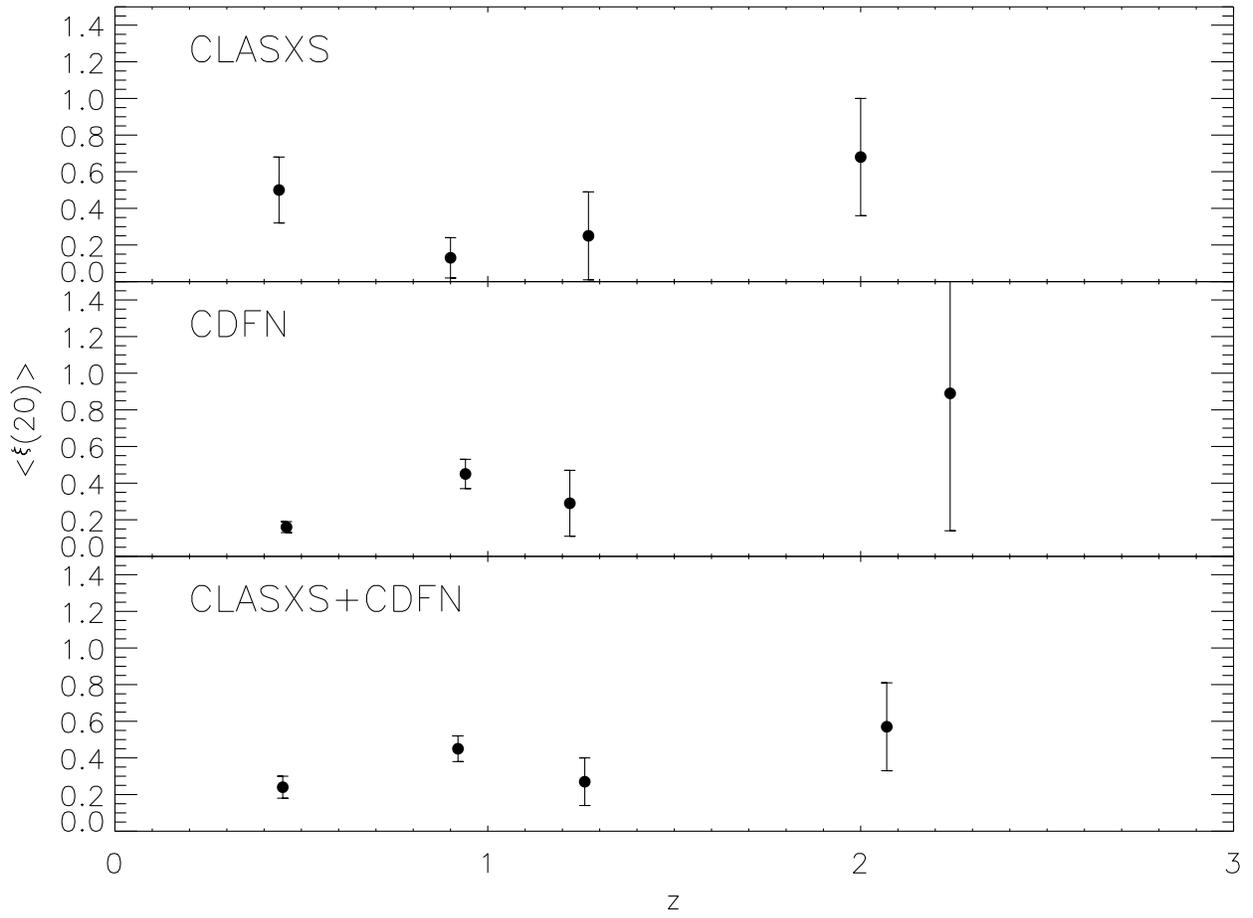}
\caption[evolution_of_xi.ps]{The evolution of clustering as 
a function of redshift for the CLASXS, CDFN and the two fields combined. 
\label{evolution3}}
\end{figure}

We study the evolution of clustering in both CLASXS and CDFN
samples, using the redshift-space correlation function. The sources 
are grouped in 4 redshift intervals from 0.1 to 3. The sizes of the 
intervals are chosen so that the number of objects in each interval 
is similar in the CLASXS sample. This result in a very wide redshift 
bin above $z=1.5$. The correlation functions for the CLASXS, CDFN and 
CLASXS+CDFN fields are shown in Figures \ref{z_dep_clasxs},
\ref{z_dep_cdfn}, and \ref{z_dep_comb}, respectively. We group
the pair separations in 10 bins in these figures to show the shape of the
correlation function. In some bins there could be no DD pairs, and 
the correlation function is set to -1 without errors. We model the correlation 
functions using single power-laws and fit the data using the maximum-likelihood 
method. As we mentioned earlier, the method is not affected by binning.  
We found on 3--50~Mpc scales that a single power-law provides a good fit to the data except, 
for the the $z=1.5-3$ interval in 
the CDFN, where the sample is too sparse and have very few close separation
pairs, we use a separation range of 5--200~Mpc to obtain 
the fit. The goodness-of-fit is quantified with $\chi^2$. In the case where 
empty bins exist, we increase the bin sizes until no bins are empty before we
compute the $\chi^2$.   
The results are summarized in Table~\ref{tab_evolution} and the $\bar{\xi}(20)$s
as a function of redshift are shown in Figure~\ref{evolution3}. 
We have tested fitting the correlation functions over different scale
ranges, and found no significant difference in the resulting $\bar{\xi}(20)$s. 

There is only mild evolution seen in both the CLASXS and CDFN
fields, in agreement with the assumption that clustering is close to constant
in comoving coordinates. There are some small discrepancies between the CLASXS 
and the CDFN clustering strength. These discrepancies give the sense of the field-to-field
uncertainty and could be resolved with future larger surveys.  At the highest redshift, 
both samples show an increase trend of clustering, but only at the $\le 2 \sigma $ level.  

\subsection{Comparing with other observations}
\begin{figure}
\includegraphics[scale=.7,angle=90]{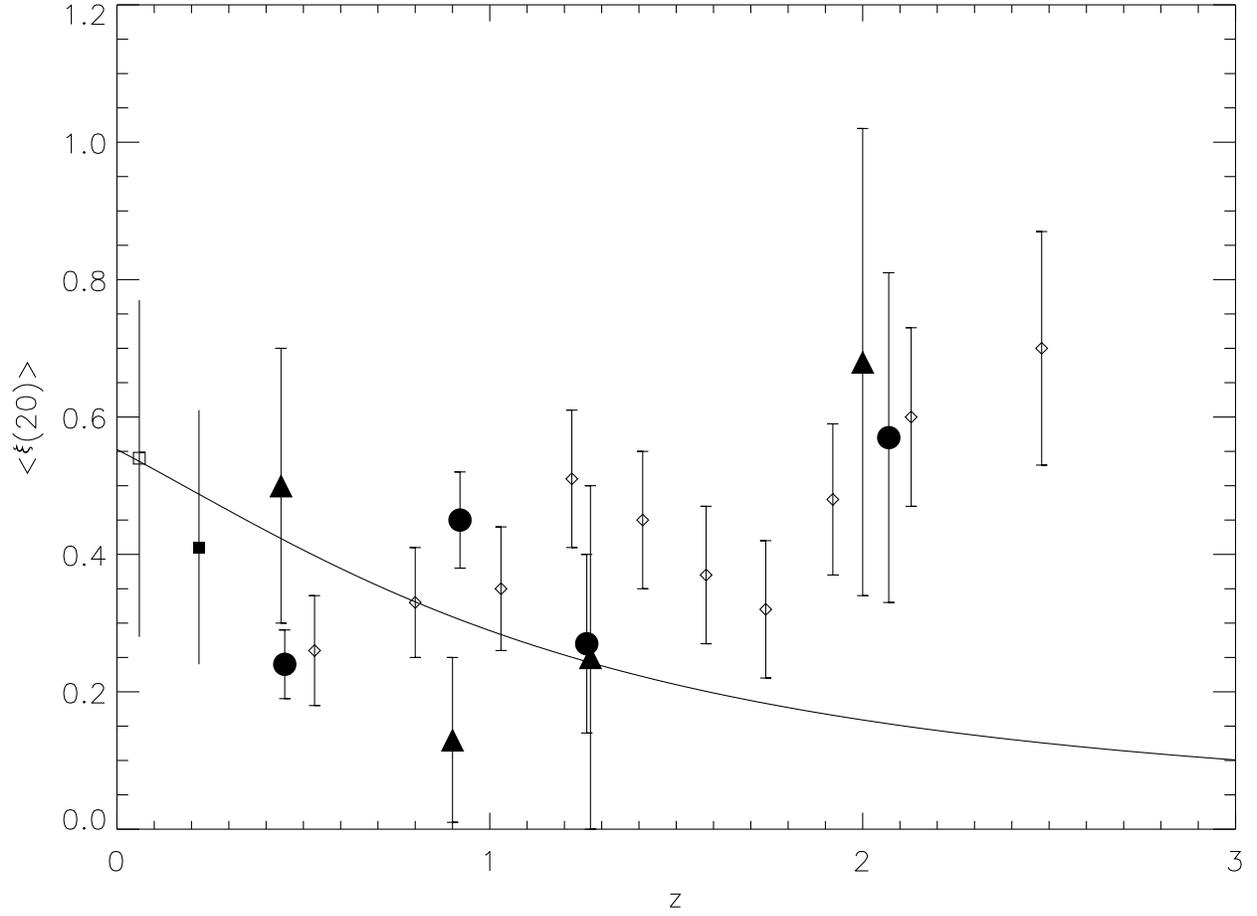}
\caption[evolution.ps]{ A comparison of clustering evolution
in the combined {\it Chandra} fields (big dots), CLASXS field (big filled
triangle), 2dF (diamonds),  ROSAT NGP (filled box) and AERQS (empty box).
The solid line represent linear evolution of clustering normalized to the
AERQS.  
\label{evolution_compare}}
\end{figure}

In Figure~\ref{evolution_compare} we plot $\bar{\xi}(20)$ as a function of redshift for 
CLASXS, the combined CLASXS and CDFN,  
as well as results from the 2dF (C05), the {\it ROSAT} North Galactic Pole Survey
(NGP, {Mullis} {et~al.} 2004), and the Asiago-ESO/RASS QSO survey (AERQS, 
{Grazian} {et~al.} 2004).  
We did not correct for redshift distortion for observations which uses redshift-space
correlation function. This leads to overestimates of the real-space correlation amplitude.
  
\begin{figure}
\includegraphics[scale=.7,angle=90]{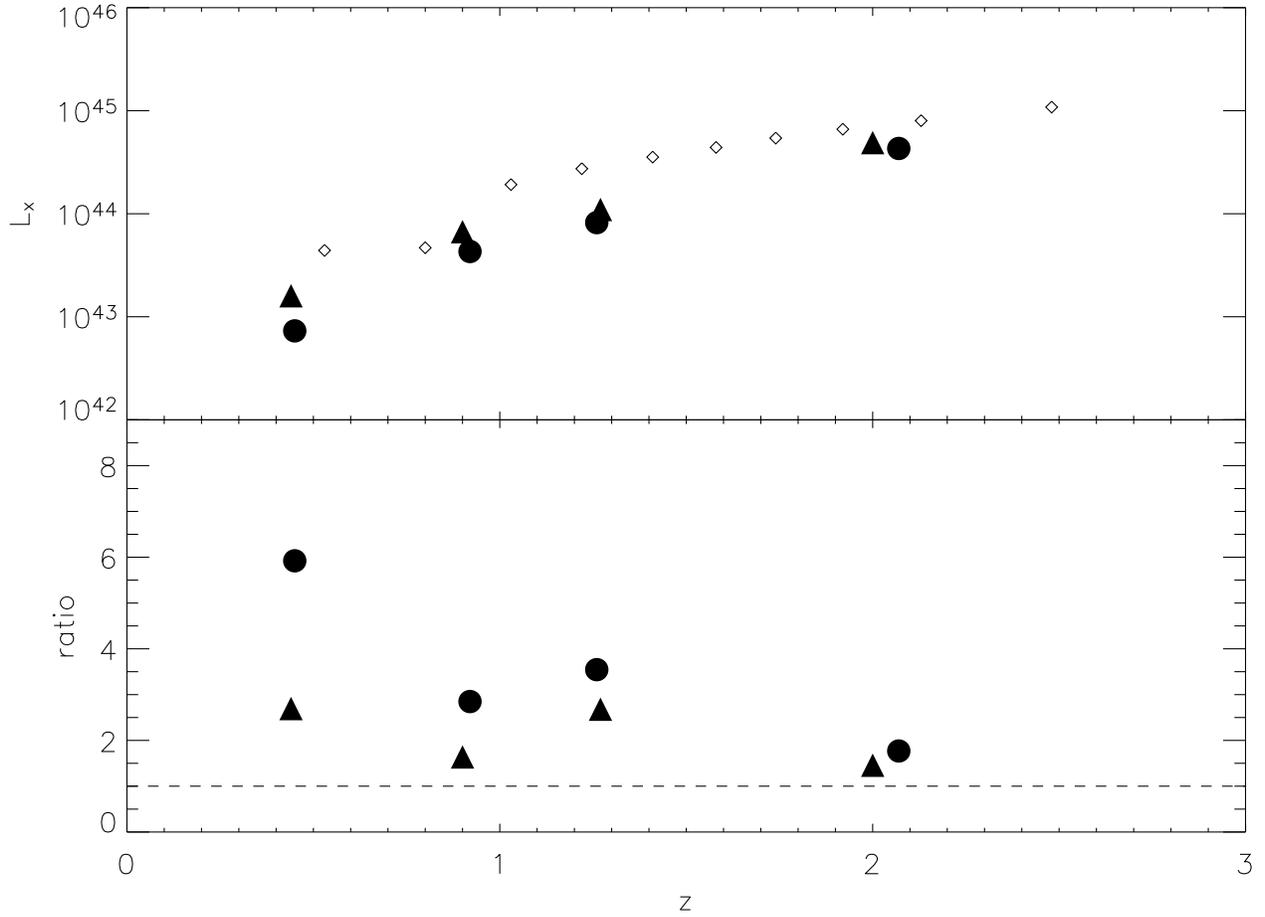}
\caption[croom_Lx_vs_z.ps]{The median luminosities of the 2dF 
quasar (C05) as a function of redshift (diamonds) compared to
the median luminosities of CLASXS sample (triangles) and of CLASXS+CDFN
sample (big dots). The lower panel shows the ratio of 2dF median luminosities 
to  the X-ray samples. 
\label{lum_comp}}
\end{figure}

Our correlation functions show clear agreement with the evolution trend found in C05.  
However, as seen in \S~\ref{sec_L_dependence}, our measured correlation amplitude on 
average appears similar as that of 2dF, even though the average
luminosity of latter is much higher than that of the X-ray samples. 
We compare the X-ray luminosities of the CLASXS and CLASXS+CDFN samples with those of the 
2dF in Figure~\ref{lum_comp}. The X-ray luminosities of 2dF quasars are obtained the same way 
as in \S~\ref{sec_L_dependence}. The luminosity difference between the 2dF sample and X-ray 
samples is the largest at low redshift and decreases at higher redshift. At $z > 2$, the X-ray sample 
and the 2dF samples have similar median luminosity. The similarity in clustering amplitude
can be understood in the light of the weak correlation between AGN luminosity and 
clustering amplitude found in \S~\ref{sec_L_dependence}. However, 
as we will see in \S~\ref{sec_xi_Lx}, a correlation between dark halo mass and the X-ray 
luminosity predicts a rapid increase of correlation function above $L_x \sim 10^{44}$~\ergps. 
Therefore, we should expect the optical sample being more clustered than X-ray samples at medium 
redshifts. Such a trend is not clearly seen in our samples.     
    
\section{Discussion
\label{sec_discussion}}
\subsection{Evolution of Bias and the typical dark matter halo mass
\label{sec_bias_mass}}
The bias evolution of optical quasar is extensively discussed in C05. They found that the
bias increases rapidly with redshift($b \sim (1+z)^2$). We will 
follow these arguments to estimate the bias evolution of the X-ray samples. 
 
\begin{figure}
\includegraphics[scale=.7,angle=90]{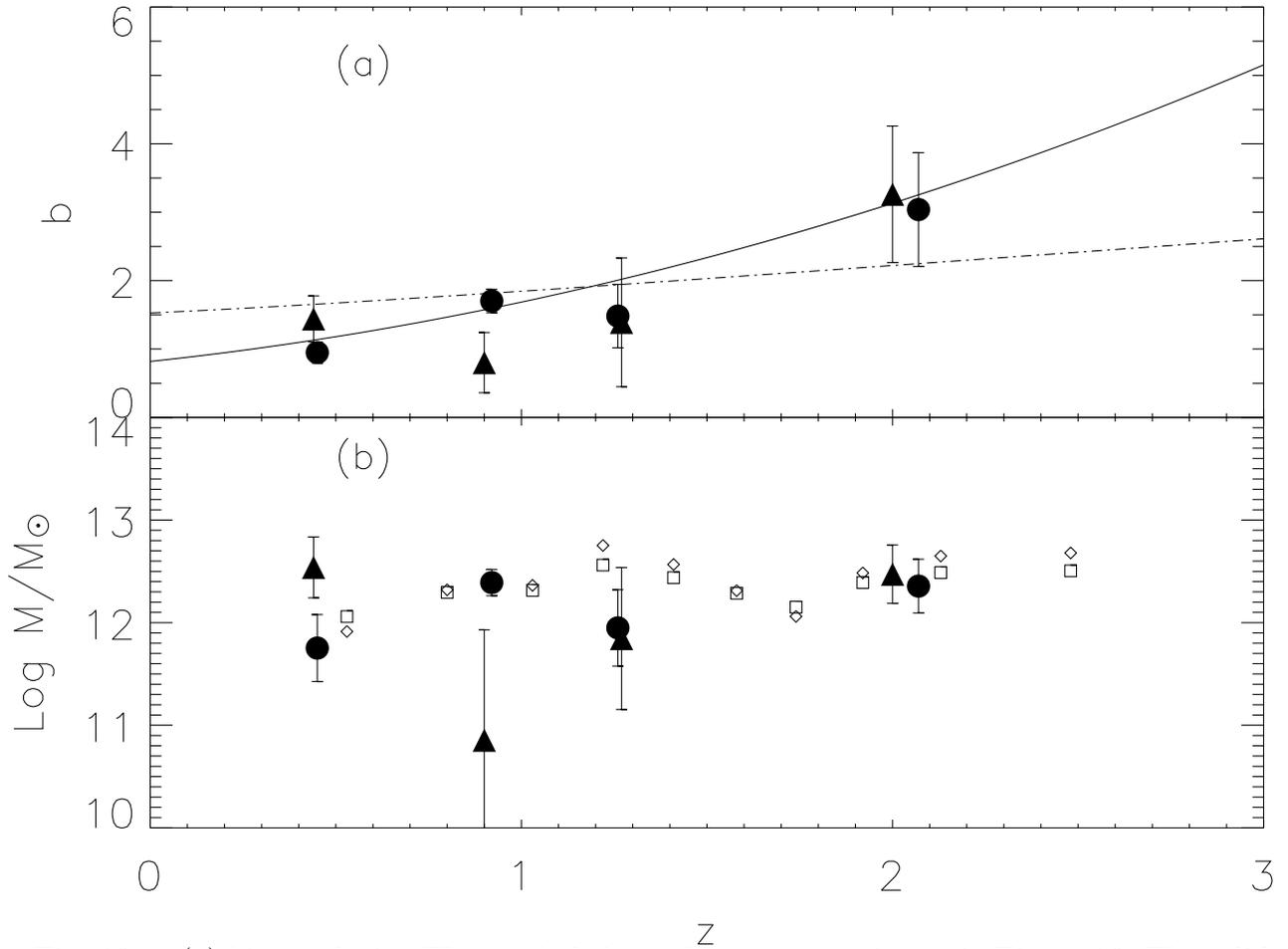}
\caption[bias_mass.ps]{(a) bias evolution.The symbols have the same meaning 
as in Figure \ref{evolution_compare}. The solid line is the best-fit from 
C05. Dash-dotted line shows the linear bias evolution model. 
(b). The  mass of host halo of the X-ray sources corresponding to the bias
in panel (a). 
\label{bias_mass}}
\end{figure}

On scales of 20~Mpc, the clustering of dark matter and AGNs are both in the 
linear regime, i.e., $\bar{\xi}(20) < 1$. This allows us to measure the bias, defined as 
the ratio of rms fluctuation of the luminous matter (AGNs in our case) and that of the underlying
total mass, as a function of redshift by comparing 
the observed correlation function with the linear growth rate. 
In terms of correlation function, the bias can be written as
\begin{equation}
b^2 \equiv \xi_{light}/\xi_{mass}.
\label{biasdef}
\end{equation}
The averaged correlation function of mass can be obtained using 
\begin{equation}
\bar{\xi}_{mass}(20) = \frac{3}{(3-\gamma)J_2(\gamma)}(\frac{8}{20})^{\gamma}\sigma_8^2 D(z)^2
\label{sigma8}
\end{equation}
where $J_2(\gamma) =72/[(3-\gamma)(4-\gamma)(6-\gamma)2^{\gamma}]$, $\sigma_8=0.84$ is the rms
fluctuation of mass at $z=0$ obtained by {\it WMAP} observation ({Spergel} {et~al.} 2003), and we choose the 
best-fit $\gamma \sim 1.5$. $D(z)$ is the linear growth factor, for which we use the approximation formula
from {Carroll}, {Press}, \& {Turner} (1992). The redshift-space distortion is taken into account to the first order through Equation 
\ref{distortion} and the bias factor is solved numerically. The result is shown in Table~\ref{tab_bias}. 
The estimate of $b(z = 1) \sim 2.2$ in the combined sample agrees with the result from the redshift-space 
distortion analysis in \S~\ref{sec_z_distortion}. In Figure~\ref{bias_mass}(a) we show the bias estimates for the CDFN and 
CLASXS+CDFN samples. The best-fit model from C05 qualitatively agrees with the X-ray results.  

The simplest model for bias evolution is that the AGNs are formed at high redshift, 
and evolve according to the continuity equation ({Nusser} \& {Davis} 1994; {Fry} 1996).
The model is sometimes called the {\it conserving model} or the {\it test particle model}.
By normalizing the bias to $z=0$, the model can be written as
\begin{equation}
b(z) = 1+(b_0-1)/D(z).
\label{bias_linear}
\end{equation}
This model is shown in Figure~\ref{bias_mass}(a) as dash-dotted line. The model 
produces a bias evolution which is slightly too shallow at high redshifts. The correlation function evolution 
based on this model is also shown in Figure~\ref{evolution_compare}, where it underpredicts the observed $\xi$. 
The model predicts a decrease of correlation function at high redshift, which is not true based on our results
and that of the 2dF. This implies that the AGNs observed in the local universe are unlikely to
have formed at $z \gg 2$.    

One of the direct predictions of CDM structure formation scenario is that the bias is determined by the dark halo mass. 
{Mo} \& {White} (1996) found a simple relation between the minimum mass of the dark matter halo and the bias $b$.
By adopting the more general formalism by {Sheth}, {Mo}, \& {Tormen} (2001) we can compute the ``typical'' dark halo mass
of the sample. It should be noted that the method assumes that halos are formed through violent collapse 
or mergers of smaller halos and hence is best applied at large separations,
where the halo-halo term dominates the correlation function. This requirement is 
apparently satisfied by AGNs. Following {Sheth}, {Mo}, \& {Tormen} (2001),
\begin{equation}
b(M,z) = 1+\frac{1}{\sqrt{a}\delta_c(z)}[a\nu^2\sqrt{a}+0.5\sqrt{a}(a\nu^2)^{(1-c)} 
-\frac{(a\nu^2)^c}{(a\nu^2)^c+0.5(1-c)(1-c/2)}],
\label{smt}
\end{equation}
where $\nu \equiv \delta_c(z)/\sigma(M,z)$, a = 0.707, c = 0.6. $\delta_c$ is the 
critical overdensity. $\sigma(M,z)$ is the rms density fluctuation in the linear 
density field and evolves as
\begin{equation}
\sigma(M,z) = \sigma_0(M) D(z), 
\label{sigma}  
\end{equation}
where $\sigma_0(M)$ can be obtained from the power spectrum of density perturbation $P(k)$
convolved with a top-hat window function $W(k)$, 
\begin{equation}
\sigma_0(M) = \frac{1}{2\pi^2} \int{dk k^2 P(k) |W(k)|^2}
\label{sigma0}
\end{equation}
At the scale of interest ($\sim 10$~Mpc), the power spectrum can be approximated with a 
power-law, $P(k) \propto k^n$, with $-2 \lesssim n \lesssim -1$ for CDM type spectrum. 
Integrating Equation~\ref{sigma0} gives
\begin{equation}
\sigma_0(M) = \sigma_8 (\frac{M}{M_8})^{-(n+3)/6},
\label{sigma0_M}
\end{equation}
where $M_8$ is the mean mass within $8~h^{-1}$~Mpc. 

We can then solve Equation~\ref{smt} for halo mass.
The resulting mass is shown in Table~\ref{tab_bias} and Figure~\ref{bias_mass}(b). Consistent with what's 
been found in C05 for the 2dF, the halo mass does not show any evolution trend with redshift. We found 
$ <\log (M_{halo}/M_{\sun})> \sim 12.11 \pm 0.29$, which is consistent with the 2dF estimates (C05, {Grazian} {et~al.} 2004).

\subsection{Linking X-ray luminosity and clustering of AGNs
\label{sec_xi_Lx}}
We have shown that over a very wide range of luminosity, the clustering amplitude 
of AGNs change very little. This allows us to put useful constrains on the correlations
among X-ray luminosity, blackhole mass $M_{BH}$, and the dark matter halo $M_{halo}$. 

Using the equivalent width of broad emission lines as mass estimators, {Barger} {et~al.} (2005) 
found that the Eddington ratio of broadline AGNs may be close to constant. 
Since the hard X-ray luminosity is an isotropic indicator of the bolometric luminosity, 
this implies that the blackhole mass is linearly correlated with X-ray luminosity.
Barger et al. (2005) found that 
\begin{equation}
{L_{44}} =(\frac{M_{BH}}{10^{8} \Msun}), 
\label{Lx_Mbh}
\end{equation} 
where $L_{44}$ is $L_x$ in units of $10^{44}$~\ergps. A similar result is 
found at low redshift using a sample of  broadline AGNs with mass estimates 
based on reverberation mapping relations ({Kaspi} {et~al.} 2000; {Yang} 2005).  
The relation, however, is only tested for broadline AGNs.  
Deviations from this relation is also expected at low luminosities since many 
low luminosity AGNs tend to have a low Eddington ratio ({Ho} 2005).

Blackhole mass have been shown to correlate with velocity dispersion of the 
spheroidal component of the host galaxies ({Gebhardt} {et~al.} 2000; {Ferrarese} \& {Merritt} 2000).
This leads to a linear correlation between $M_{BH}$ and the mass of spherical
component. This relation, however, could be different at high redshift 
({Akiyama} 2005). How these relationships translate to the $M_{BH} - M_{halo}$ 
relation is also unclear and could likely be nonlinear. {Ferrarese} (2002) 
showed that $M_{BH}$ -- $M_{halo}$ can be modeled with a  scaling law      
\begin{equation}
\frac{M_{BH}}{10^8 \Msun} = \kappa(\frac{M_{halo}}{10^{12} \Msun})^{\lambda}, 
\label{BH_Halo}
\end{equation}
with $\kappa$ and $\lambda$ determined by the halo mass profile.  

Combining the above and using  Equation~\ref{smt}, we can calculate the correlation amplitude as a 
function of X-ray luminosity. In Figure~\ref{Lum_depend} we show the model expectations compared 
with the observations from CLASXS, CDFN and 2dF. In calculating the bias we have assumed the 
nonlinear power-law index $n = 3-\gamma$  in Equation~\ref{sigma0_M} (Peacock 1999), with the best 
fit $\gamma=1.5$.  The three lines represent three different halo profiles discussed
in {Ferrarese} (2002). We found that the $L_{x} - \bar{\xi(20)}$ relation 
is in fact dominated by the very nonlinear relation between halo mass and 
correlation amplitude. The difference between different halo profiles
is caused mainly by the normalization $\kappa$, or the fractional mass of blackhole mass, rather than 
the power-law index $\lambda$. One of the important model predictions is that the correlation between
X-ray luminosity and clustering is weak below $\sim 10^{43}$~\ergps and increases
rapidly above that. The lack of rapid change of correlation amplitude indicates
the halo mass of AGN cannot be significantly higher than the corresponding threshold. 
Under the assumed cosmology and bias model, the $L_{x} - \bar{\xi(20)}$
relation based on the weak lensing derived halo mass profile ({Seljak} 2002)
is consistent with the data, while the NFW profile 
({Navarro}, {Frenk}, \& {White} 1997)  and  the isothermal profile predicts a too steep 
correlation amplitude curve at high luminosity. However, we cannot rule out these profiles as a reasonable descriptions of the AGN host halo because of the uncertainty in the 
shape of the correlation function and the fractional blackhole mass in dark halos at the redshift of our sample. In Figure~\ref{Lum_depend} we also mark the model dark  
halo mass corresponding to the Seljak (2002) mass profile. The average correlation amplitude of the 
combined optical and X-ray sample (dotted-line) corresponds to a halo mass of $\sim 2 \times 10^{12}$~\Msun. 
While the luminosity in our sample ranges over five orders of magnitudes, the range of halo mass may be 
much smaller. The 2dF sample has a 
high luminosity but has a similar or lower average correlation amplitude than that of the 
X-ray samples. It is possible that the optical selection
technique tends to select sources with a higher Eddington ratio.  
The correlation amplitude of the CDFN sample at $\sim 10^{41}$~\ergps, on the other hand,
is higher than the model predictions. This is expected because many AGNs with 
such luminosities are LINERs which are probably accreting with a low radiative
efficiency. 

It is now clear that the weak luminosity dependence of AGN clustering is consistent
with the simplest model based on the observed $L_x - M_{BH}$ and $M_{BH} - M_{halo}$ 
relations, given the large error in the correlation functions. A large dynamical range in X-ray
luminosity, as well as better measurements of correlation function, are needed to better quantify this 
relation. The luminosity range of the 2dF survey is too small and the optical selection method also likely 
is biased to high Eddington ratio sources. By increasing our current CLASXS field by a factor of a few will be 
helpful in better determine the luminosity dependence of AGN clustering, and to put tighter constrains 
on AGN hosts. 

\subsection{Blackhole mass and the X-ray luminosity evolution}
We look again at the $M_{BH}$--$M_{halo}$ relation in the light of the 
mass estimates of the dark matter halos from {\it Chandra} samples. 
If the {Ferrarese} (2002) relation is independent of redshift, the nearly constant dark halo mass
implies little evolution for the blackhole mass.
On the other hand, strong luminosity evolution is seen since $z = 1.2$ in hard X-ray selected AGNs
({Barger} {et~al.} 2005). This implies a systematic decrease of the ensemble Eddington ratio with cosmic time. 
{Barger} {et~al.} (2005) showed that the characteristic luminosity of hard X-ray selected AGNs
\begin{equation}
L^{\star} = L_0 (\frac{1+z}{2})^a, 
\label{L_star}
\end{equation}
where $\log (L_0/\ergps) = 44.11$ and $a = 3.2$ for $z<1.2$. 
{Ueda} {et~al.} (2003) found a similar result with a slightly shallower slope. 
If the typical blackhole mass does not change with redshift, the observed luminosity evolution 
indicates the ensemble Eddington ratio increase by a factor of $\sim 10$ from $z=0$ to $z=1$.
It is hard to understand such a change of the typical Eddington ratio with redshift.  
One possibility is that a large number of Compton thick 
AGNs at $z \sim 1$ are missed in the {\it Chandra} surveys (e.g. {Worsley} {et~al.} 2005), 
leading to the observed strong luminosity evolution. 

\begin{figure}
\includegraphics[scale=.7,angle=90]{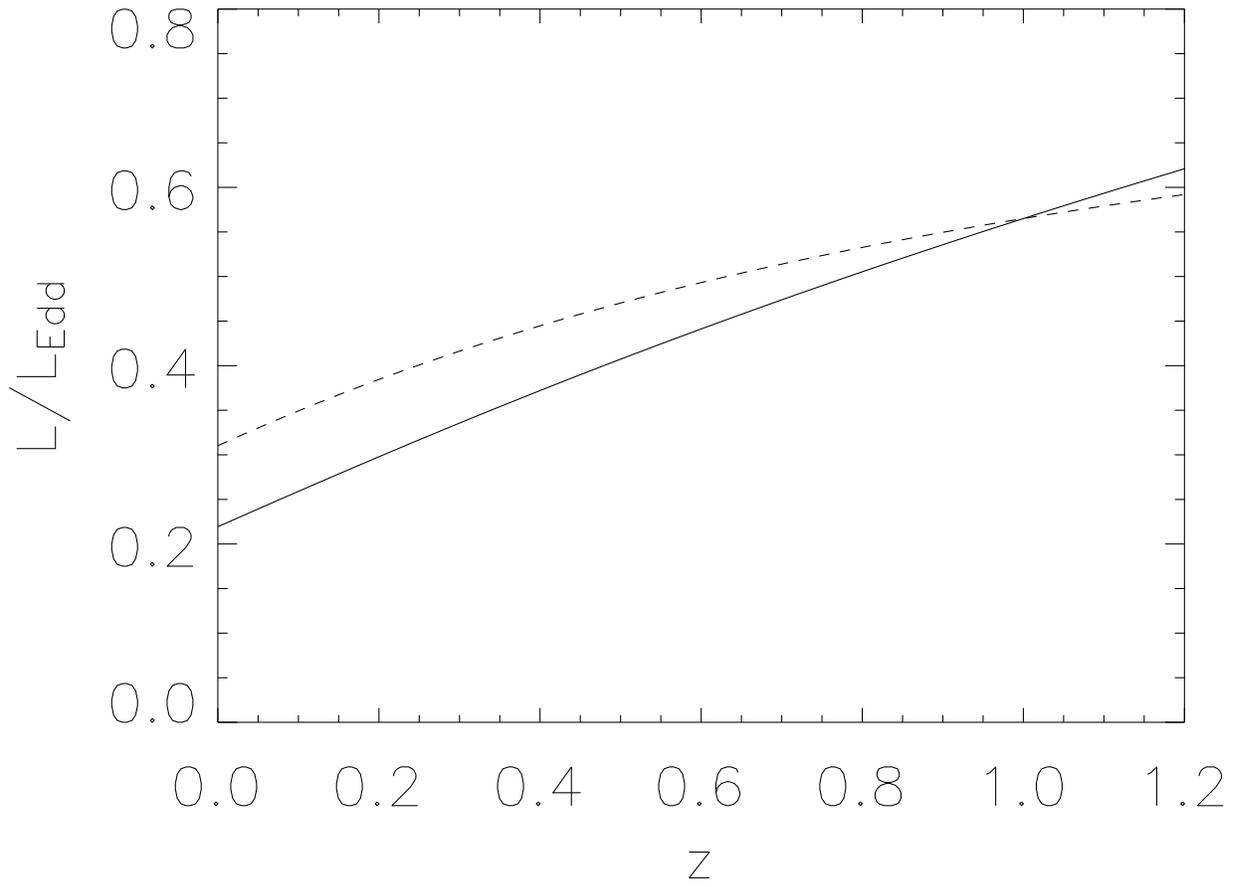}
\caption[Eddington_ratio.ps]{ Evolution of Eddington ratio. Solid line:
Using luminosity function from Barger et al. (2005). Dashed line:
using luminosity function from \citet{ueda03} at $z< 1.2$.  
\label{eddington}}
\end{figure}

Alternatively, instead of $M_{BH}$--$M_{halo}$ being independent of redshift, the
$M_{BH}$--$v_c$ could be unchanged with cosmic time, as suggested by {Shields} {et~al.} (2003).
This is theoretically attractive because the feedback regulated growth of blackholes
implies a constant $M_{BH}$--$v_c$ relation. This implies that  $M_{BH}$--$M_{halo}$ is in 
fact a function of redshift. {Wyithe} \& {Loeb} (2003) proposed a model(WL model here after) showing that  
the blackhole mass inferred from the halo mass increases with redshift. {Croom} {et~al.} (2005) show that 
this could lead to a close to constant Eddington ratio in the 2dF sample if the optical luminosity 
is used to compare with the derived $M_{BH}$. Since the correlation function is only a weak function
of luminosity, as we have demonstrated in \S~\ref{sec_L_dependence}, it is better to estimate the evolution of 
the Eddington ratio using the characteristic mass of the blackholes from the WL model, 
and the characteristic luminosity from Equation~\ref{L_star}. In Figure~\ref{eddington}, we show the 
derived ensemble Eddington ratio, assuming the dark halo mass to be constant and 
$\log~(<M_{halo}/M_{\sun}>) \sim 12.11$.
(we adopt the normalization of the WL model so that it matches the prediction
of $M_{BH}$ -- $M_{halo}$ with a NWF type of halo profile. However, the choice
of this normalization is not crucial). In the figure, we see a factor of $\sim 2-3$ change in the ensemble Eddington
ratio from $z=0$ to $z = 1.2$. This change, however, is smaller than the typical scatter in both the luminosity 
and halo mass. 
  
\subsection{Comparison with normal galaxies}
We now compare our clustering results with those for normal galaxies. 
Using the Sloan Digital Sky Survey First Data Release,  
{Wake} {et~al.} (2004) found that the 
clustering of narrow-line AGNs in the redshift range $0.055 < z < 0.2$, 
selected using emission-line flux ratios, have the same correlation amplitude 
as  normal galaxies. Our samples are not a very good probe at these
redshifts, and the best clustering analysis at a comparable
redshift for normal galaxies is from DEEP2 ({Coil} {et~al.} 2004). At effective 
redshift $z_{eff} \sim 1$, they found $r_0 = 3.19 \pm 0.51$~h$^{-1}$~Mpc,
and $\gamma = 1.68 \pm 0.07$, which translates to $\bar{\xi}(20) \sim 0.1$. 
The correlation amplitude from CLASXS at $z = 0.9$ is $\bar{\xi}(20) \sim 0.13^{+0.12}_{-0.10}$.
The clustering of AGNs in CLASXS field appear the same as the clustering 
of normal galaxies in DEEP2.  A higher  correlation is found in the CDFN. The difference shows the large uncertainty of our correlation function 
estimates.  At higher redshifts, the best estimate for galaxy clustering is from the so called ``Lyman break galaxies'', 
named after the technique by which they are  found. {Adelberger} {et~al.} (1998) found, at a typical $z \sim 3$,
these galaxies tend to have similar correlation function as galaxies in the local universe, 
indicating they are highly biased tracers of the large scale structure. In the 
$\Lambda$CDM cosmology, these authors found $b = 4.0 \pm 0.7$. This is very similar
to the bias found in the highest bin of our {\it Chandra} fields ($b = 3.03 \pm 0.83$), which 
has a median redshift of $\sim 2.0$. If we extrapolate the bias of the X-ray sources
to $z=3$, the bias of X-ray sources should be $\sim 4-5$, consistent with the clustering strength
of Lymann break galaxies.      

\section{Conclusion
\label{sec_conclusion}}
In this paper we study the clustering of the X-ray selected AGNs in the 0.4~deg$^{2}$ {\it Chandra} 
contiguous survey of the Lockman Hole Northwest region. Based on our previous
study, the size of the CLASXS field is large enough that the cosmic variance should not
be important. We supplement our study with the published data from the CDFN.  
The very similar LogN-LogS of the CLASXS and CDFN suggests that the 
cosmic variance should not be important when the CDFN is included in the analysis. 
The very deep CDFN gives a better probe of the correlation function at small separations. 
A total of 233 and 252 non-stellar sources from CLASXS and CDFN respectively are used in this study. 
We use the correlation function in the redshift-space as a major tool in our clustering analysis. 
For the whole sample, we have also performed an analysis using the
projected correlation function. This allows us to quantify the effect of redshift distortion.   

We summarize our results as follows:

1. We calculated the redshift-space correlation function for sources
with $0.1<z<3.0$ in both the CLASXS and CDFN fields, assuming constant clustering in 
comoving coordinates. We found a $6.7\sigma$ clustering signal for pairs within $s<20$~Mpc
in the CLASXS field. The correlation function over scale of 3~Mpc$<s<$~200~Mpc is found 
to be a power-law with $\gamma = 1.6^{+0.4}_{-0.3}$ and $s_0 = 8.0^{+1.4}_{-1.5}$~Mpc. 
The redshift-space correlation function for CDFN on scales of 1~Mpc$<s<$~100~Mpc is 
found to have similar correlation length $s_0 = 8.55^{+0.75}_{-0.74}$~Mpc, 
but the slope is shallower ($\gamma = 1.3 \pm 0.1$). 

2. We study the projected correlation function of both CLASXS and CDFN. 
The best-fit parameters for the real-space correlation functions are found to be
$r_0 = 8.1^{+1.2}_{-2.2}$~Mpc, $\gamma = 2.1 \pm 0.5$ for CLASXS field, 
and $r_0 = 5.8^{+1.0}_{-1.5}$~Mpc, $\gamma = 1.38^{+0.12}_{-0.14}$ for CDFN field. 
Our result for the CDFN shows perfect agreement with the published results from Gilli et al. (2004).
Fitting the combined data from both fields gives  $r_0 = 6.1^{+0.4}_{-1.0}$~Mpc and
$\gamma = 1.47^{+0.07}_{-0.10}$. 

3. Comparing the redshift- and real-space correlation function of the combined CLASXS and CDFN fields, 
we found the redshift distortion parameter $\beta = 0.4 \pm 0.2$ at an effective redshift $z = 0.94$. 
Under the assumption of $\Lambda$CDM cosmology, this implies a bias parameter  $b \approx 2.04 \pm 1.02$
at this redshift.

4. We tested whether the clustering of the X-ray sources is dependent on the X-ray spectra in the CLASXS field. 
Using a hardness ratio cut at $HR = 0.7$, we found no significant difference in clustering between hard and 
soft sources. This agrees with previous claims. 

5. With the large dynamic range in X-ray luminosity, we found  a  weak correlation between 
X-ray luminosity and clustering amplitude. Using a simple model based on observations that links
the AGN luminosity and halo mass, we show that the observed weak correlation is consistent with 
the model, except at low luminosities, where sources are likely to have lower Eddington ratio. 
The non-detection of a strong correlation between X-ray luminosity and clustering amplitude
also suggests a narrow range of halo mass. 

6. We study the evolution of the clustering using the redshift-space correlation function in 4 redshift intervals
ranging from 0.1 to 3.0. We found only a mild evolution of AGN clustering in both CLASXS and CDFN samples.
This qualitatively agrees with the results based on optically selected quasars from 2dF survey. The X-ray
samples, however, show a similar correlation amplitude as that of the 2dF sample. This is 
consistent with the weak correlation between AGN luminosity and the clustering amplitude
found in this work.  
 
7. We estimate the evolution of bias by comparing the observed clustering amplitude with expectations of 
the linear evolution of density fluctuations. The result show that the bias increases rapidly with redshift 
($b(z=0.45) \sim 0.95$ and $b(z = 2.07) \sim 3.03$ in CLASXS field). This agrees with the findings from 2dF. 

8. Using the bias evolution model for dark halos from Sheth, Mo \& Tormen (2001), we estimated the 
 characteristic mass of AGNs in each redshift interval. We found the mass of the dark halo changes very 
little with redshift. The average  halo mass is found to be  $\log~(M_{halo}/M_{\sun}) \sim 12.11$.

Our results have demonstrated that deep X-ray surveys are a very useful tool in studying how AGNs trace the 
large scale structure. Such knowledge provides an unique window to the understanding of  
AGN activity, and its relation to structure formation. The higher spatial density and much better 
completeness compared to current optical surveys allows us to study clustering on scales only accessible 
to very large optical surveys such as the 2dF and the SDSS. The high spatial resolution and positional 
accuracy of {\it Chandra} is critical for unambiguous optical identifications. Since our results on the evolution of AGN clustering 
could still be affected by a small number of large scale structures, as seen in Chandra Deep Field South, larger 
{\it Chandra} fields are still needed to improve the measurements.     

\acknowledgements
We thank Prof. Alex Szalay,  Dr. Xaviar Barcons and Dr. Takamitsu
Miyaji for very helpful conversations;  Dr. Chris Mullis for
discussions on his paper. We would like to thank the anonymous referee
for his/her comments that help to improve the paper. This work is
partially funded by the IDS program of R. F. M. We also gratefully 
acknowledge support from NSF grants AST 02-39425 (A.J.B.) and 
AST 04-07374 (L.L.C.), the University of Wisconsin Research Committee
with funds granted by the Wisconsin Alumni Research Foundation,
the Alfred P. Sloan Foundation, and the David and Lucile Packard
Foundation (A.J.B.).


\clearpage
\begin{deluxetable}{lccclccc}
\tabletypesize{\footnotesize}
\tablewidth{0pt}
\tablecaption{Redshift-space Correlation Function}
\tablecolumns{8} 
\tablehead{ 
\multicolumn{4}{c}{CLASXS Field} &
\multicolumn{4}{c}{ CDF-N Field} \\
\hline
\colhead{s range (Mpc)} &
\colhead{$s_0$} &
\colhead{$\gamma$} & 
\colhead{$\chi^2/dof$} &
\colhead{s range (Mpc)} &
\colhead{$s_{0}$} & 
\colhead{$\gamma$} &
\colhead{$\chi^2/dof$}
} 
\startdata 
10--200 & $11.4^{+1.8}_{-3.1}$ & $2.4^{+1.1}_{-0.8}$ & 6.2/8 &
10--100 & $11.5^{+0.8}_{-1.2}$ & $2.9^{+1.4}_{-0.8}$ &  7.9/8 \\
 3--30 & $8.15^{+1.6}_{-2.0}$ & $1.2^{+0.5}_{-0.4}$ & 3.8 /8 & 
 1--20 & $11.4^{+1.8}_{-1.4}$ & $.96^{+.15}_{-.17}$ &  6.8/8 \\
 3--200 & $8.05^{+1.4}_{-1.5}$ & $1.6^{+0.4}_{-0.3}$ & 10.6/8 &
 1--100 & $8.55^{+.75}_{-.74}$ & $1.3 \pm 0.1$       & 15.0/8 \\ 
\enddata 
\label{z_space_cor}
\end{deluxetable} 
\vfil\eject\clearpage
\begin{deluxetable}{cccccccc}
\tabletypesize{\footnotesize}
\tablewidth{0pt}
\tablecaption{Luminosity dependance of Correlation Function}
\tablecolumns{8} 
\tablehead{ 
\colhead{Field} &
\colhead{z range} & 
\colhead{z$_{median}$} & 
\colhead{$<L_{x}>$ (\ergps)} & 
\colhead{$s_0$} &
\colhead{$\gamma$} & 
\colhead{$\chi^2/dof$} &
\colhead{$\bar{\xi}(20)$} 
} 
\startdata 
CLASXS & 0.1--3.0 & 1.5 & $3.3 \times 10^{44}$ & 
$11.5^{+1.9}_{-2.1}$ & $2.0^{+.5}_{-0.4}$ & 7.2/8  & $0.50^{+.18}_{-.17}$\\
       & 0.1--3.0 & .73 & $1.5 \times 10^{43}$ & 
$7.35^{+1.9}_{-2.0}$ & $1.9^{+1.2}_{.54}$ & 8.8/8 &  $.21^{+17}_{-.11}$ \\
       & 0.3--1.5 & 1.1 & $1.4 \times 10^{44}$ &  
$11.0 \pm 2.6$      & $2.3^{+1.6}_{-0.6}$ & 9.2/8 & $0.49^{+.31}_{-.23}$ \\
       & 0.3--1.5 & .81 & $1.6 \times 10^{43}$ & $5.30^{+2.9}_{-3.8}$ & 
$1.4^{+0.8}_{-0.5}$ &  7.8/8  & $.18^{+.15}_{-.14}$\\
CDF-N  & 0.1--3.0 & .98 & $7.9 \times 10^{43}$ & $13.2 \pm 2.9$  & $.81^{+0.20}_{-0.17}$ & 8.2/8 & $.74 \pm 0.13$\\
       & 0.1--3.0 & .51 & $8.3 \times 10^{41}$ & $5.6^{+1.2}_{-1.1}$ & $1.26^{+0.22}_{-0.20}$ &  11.9/8 &$.22 \pm .05$ \\
       & 0.3--1.5 & .96 & $4.0 \times 10^{43}$ &  $8.0^{+1.5}_{-1.4}$ & $1.11^{+.25}_{-.22}$ &  11.1/8 & $.39^{+.08}_{-.07}$ \\
       & 0.3--1.5 & .63 & $1.0 \times 10^{41}$ &  $6.8^{+1.3}_{-1.2}$ & $1.28^{+.27}_{-.21}$ & 8.4/8 & $.28 \pm .07$ \\
\enddata 
\label{tab_lum}
\end{deluxetable} 
\vfil\eject\clearpage
\begin{deluxetable}{lcccccccc}
\tabletypesize{\footnotesize}
\tablewidth{0pt}
\tablecaption{Evolution of redshift-space Correlation Function}
\tablecolumns{9}
\tablehead{
\colhead{Field} &
\colhead{z range} &
\colhead{$<z>$} &
\colhead{N\tablenotemark{a}} & 
\colhead{$<L_{x}>$\tablenotemark{b}} & 
\colhead{$s_0$} &
\colhead{$\gamma$} & 
\colhead{$\chi^2/dof$} &
\colhead{$\bar{\xi}(20)$} 
}
\startdata 
CLASXS & 0.1--0.7 & 0.44 & 57 &  $1.6 \times 10^{43}$ & $10.6^{+3.2}_{-3.0}$ & 
$1.3^{+0.7}_{-0.5}$ & 4.1/8 & $0.50^{+0.20}_{-0.17}$ \\ 
       & 0.7--1.1 & 0.90 & 60 &  $6.7 \times 10^{43}$ &  $6.2^{+2.1}_{-2.8}$ & 
$2.3^{+6.0}_{-1.0}$ & 5.9/8 & $0.13^{+0.12}_{-0.10}$ \\ 
       & 1.1--1.5 & 1.27 & 49 &  $1.1 \times 10^{44}$ &  $6.4^{+5.0}_{-4.6}$ & 
$1.3^{+1.2}_{-0.7}$ & 1.6/3 & $0.25^{+0.29}_{-0.20}$ \\
        & 1.5--3.0 & 2.00 & 67 &  $4.9 \times 10^{44}$ &  $13.6^{+4.2}_{-5.4}$ & 
$1.4^{+0.6}_{-0.5}$ & 3.1/3 & $0.68^{+0.31}_{-0.34}$ \\
\hline
CDFN & 0.1--0.7 & 0.46 & 111 &  $2.8 \times 10^{42}$ & $6.8^{+0.7}_{-0.6}$ & 
$2.2^{+0.5}_{-0.3}$ & 12.5/8 & $0.16^{+0.04}_{-0.03}$ \\ 
       & 0.7--1.1 & 0.94 & 91 &  $2.6 \times 10^{43}$ &  $9.4^{+1.3}_{-1.4}$ & 
$1.2^{+0.3}_{-0.2}$ & 5.6/8 & $0.45 \pm 0.08$ \\ 
       & 1.1--1.5 & 1.22 & 28 &  $3.8 \times 10^{43}$ &  $8.8^{+2.6}_{-2.3}$ & 
$2.1^{+1.0}_{-0.8}$ & 2.9/8 & $0.29^{+0.21}_{-0.14}$ \\
        & 1.5--3.0 & 2.24 & 22 &  $2.4 \times 10^{44}$ &  $14.2^{+8.5}_{-7.9}$ & 
$2.3^{+2.2}_{-1.4}$ & 1.4/7 & $0.89^{+1.72}_{-0.75}$ \\
\hline
CLASXS+CDFN & 0.1--0.7 & 0.45 & 168 &  $7.3 \times 10^{42}$ & $7.9^{+0.9}_{-0.9}$ & 
$1.9^{+0.3}_{-0.3}$ & 5.3/8 & $0.24^{+0.06}_{-0.05}$ \\ 
       & 0.7--1.1 & 0.92 & 151 &  $4.3 \times 10^{43}$ &  $10.1^{+1.1}_{-1.0}$ & 
$1.4^{+0.2}_{-0.2}$ & 5.5/8 & $0.45^{+0.07}_{-0.06}$ \\ 
       & 1.1--1.5 & 1.26 & 77 &  $8.2 \times 10^{43}$ &  $8.4^{+1.8}_{-2.4}$ & 
$2.0^{+0.8}_{-0.6}$ & 1.8/8 & $0.27 \pm 0.13$ \\
        & 1.5--3.0 & 2.07 & 89 &  $4.3 \times 10^{44}$ &  $12.4^{+2.7}_{-3.4}$ & 
$1.7^{+0.5}_{-0.4}$ & 4.2/7 & $0.57^{+0.23}_{-0.24}$ \\

\enddata
\tablenotetext{a}{The number of sources}
\tablenotetext{b}{Unit: \ergps}
\label{tab_evolution}
\end{deluxetable} 
\vfil\eject\clearpage
\begin{deluxetable}{cccccc}
\tabletypesize{\footnotesize}
\tablewidth{0pt}
\tablecaption{Bias evolution and dark matter halo mass}
\tablecolumns{6}
\tablehead{
\multicolumn{3}{c}{CLASXS}  & \multicolumn{3}{c}{CLASXS+CDFN} \\
\hline 
\colhead{$<z>$} &  
\colhead{$b$} &
\colhead{$Log_{10} (M/M_\sun)$} & 
\colhead{$<z>$} &  
\colhead{$b$} &
\colhead{$Log_{10} (M/M_\sun)$}  
}
\startdata
0.44 & $1.44 \pm 0.34$ & $12.54 \pm 0.30$ & 0.45 & $0.95 \pm 0.15$ & $11.75 \pm 0.32 $ \\
0.90 & $0.80 \pm 0.44$ & $10.85 \pm 1.07$ & 0.92 & $1.70 \pm 0.17$ & $12.39 \pm 0.13 $ \\
1.27 & $1.39 \pm 0.94$ & $11.84 \pm 0.69$ & 1.26 & $1.48 \pm 0.46$ & $11.95 \pm 0.37 $ \\
2.00 & $3.26 \pm 1.00$ & $12.47 \pm 0.28$ & 2.07 & $3.03 \pm 0.83$ & $12.35 \pm 0.26 $ \\
\enddata
\label{tab_bias}
\end{deluxetable} 

\clearpage

\clearpage

\end{document}